\documentclass[aps,preprint,onecolumn,nofootinbib]{revtex4}

\usepackage{bm}
\usepackage{mathrsfs}
\usepackage{amssymb}
\usepackage{amsmath}
\usepackage{amsfonts}
\usepackage{color}
\usepackage{ulem}

\usepackage[]{graphicx}
\begin{document}
\newcommand{\half}{\frac{1}{2}}
\title{Making sense of Born's rule $p_\alpha=\lVert\Psi_\alpha\rVert^2$ with the many-minds interpretation }
\author{Aur\'elien Drezet$^{1}$}
\address{$^1$Univ. Grenoble Alpes, CNRS, Grenoble INP, Institut Neel, F-38000 Grenoble, France}

\email{aurelien.drezet@neel.cnrs.fr}
\begin{abstract}
This work is an attempt to justify Born's rule within the framework of the many-minds interpretation seen as a development of the many-worlds interpretation of Everett. More precisely, here we develop a unitary model of many-minds based on the work of Albert and Loewer (Synthese \textbf{77}, 195 (1988)). At the difference of previous models ours is not genuinely stochastic and dualistic and also involves some classical-like randomness concerning the initial conditions of the Universe. We also compare the present method for recovering Born's rule with previous works based on decision theory \emph{\`a la }Deutsch, Wallace, and envariance \emph{\`a la} Zurek and show how these approaches are connected to our model.        
\end{abstract}

\maketitle

\section{Introduction and motivation}\label{sec1}
\indent The many-worlds interpretation (MWI) proposed by H. Everett in 1957~\cite{Everett1957,Barrett2012,DeWitt1973} is an attempt for providing a complete unitary ontology to quantum mechanics. However, the MWI has some unwarranted features concerning the role of probability plaguing attempts for recovering the standard quantum predictions (i.e., the famous Born rule). During the last two decades  new proposals have been discussed in order to make sense of quantum probability in the MWI by using subjectivist and personalist concepts such as self-locating uncertainty, degrees of belief, Laplacian principle of indifference, `envariance'. However, none of these `Bayesian' and `decision-theoretic' based scenarios have successfully and objectively recovered the Born rule, or be able to unambiguously interpret the concept of probability. Nevertheless, all these interesting ideas provide clues which strongly motivate the present work.\\
\indent  As we show in this work everything is tied to the meaning given to quantum measurements in the MWI, i.e., as experienced and memorized by the observers. One of the great idea of Everett (beside the central idea of taking seriously unitarity for the evolution of the whole Universe) was to treat the observer quantum mechanically as a memory device or automaton in order to develop a self-consistent theory of measurement within the MWI. As Everett indeed wrote in 1957: 
\begin{quote}
\textit{As models for observers we can, if we wish, consider automatically functioning machines, possessing sensory apparatus and coupled to recording devices capable of registering past sensory data and machine configurations.}~\cite{Everett1957} 
\end{quote}
 This strategy in turn had a huge impact on the interpretation of probabilities by observers located in branches of the evolving wave-function. In this work we are going to exploit further this idea of preserving unitary  for describing  the observers.  Here, we will emphasize the role of some old ideas about minds and brain elaborated in the late 1980's and  early  1990's in connection with the so-called many-minds interpretation (MMI) of Albert and Loewer \cite{AlbertLoewer1988,Albert1994} that was mainly stochastics.  Our work is also motivated by past attempts made by Lockwood~\cite{Lockwood1989,Lockwood1996a} and Donald~\cite{Donald1,Donald2,Donald3} preserving the unitarity of the quantum evolution in a sense similar to the original MWI.  Here, based on a new unitary formulation of the MMI we develop a completely self-consistent toy model for a quantum observer that recovers Born's probabilities (in that sense the MMI is indeed a particular application of the MWI solving the probability problems).   \\
\indent  The paper is structured as follow: In Sections \ref{sec2a} and  \ref{sec2b} we will review the probability conundrum in the MWI by summarizing the main past attempts for solving the problem.  We will start with the original Everett interpretation in Section \ref{sec2a} and also present the MMI of Albert and Loewer which plays a key role in our work in Section \ref{sec2b}. In Sections \ref{sec2d} we will discuss the quantitative problem of justifying Born's rule in the MWI and focus our analysis on the work made by Deutsch \cite{Deutsch1999}, Wallace \cite{Wallace2012} and Zurek \cite{Zurek2005} in the recent years using  Bayesian deductions and `envariance'. We stress that our work doesn't directly rely on the Wallace, Deutsch and Zurek deductions but that a clear description of their analysis helps the discussion and provides clues for introducing our own reasoning.  Finally, in Section \ref{sec3} we will propose a unitary toy model for minds in the context of the MWI. In turn  this unitary MMI scenario will be used for making sense of quantum probabilities. The proposal strongly differs from the previous attempts by introducing a classical-like molecular chaos hypothesis for describing the distribution of initial conditions for qubits interacting with the minds. In other words, we will show that by taking into account the problem of the mind observer (i.e., described in the context of the all-unitary MWI together with some ingredients of randomness coming from quantum entanglement with the local environment) help us to decipher the still ambiguous concept of `self-locating uncertainty' probability in the MWI~\cite{Sebens2016,McQueen} i.e., assuming our idea of a unitary version of the MMI. Therefore, we obtain a physical and dynamical model for minds justifying why our subjective notion of probability (i.e, credence) must equal the objective one associated with the gas of qubits interacting with the minds.  Ultimately, we justify and recover the Born rule and propose a physical interpretation of the probabilities present in this model.
\section{Everett and the meta-theorem: the incoherence problem} \label{sec2a}
\indent The MWI has been the subject of intense and recurrent debates concerning the meaning and role to be given to quantum probabilities in this theory. Indeed, the MWI is a purely deterministic theory admitting strict unitary Schr\"odinger evolution as the only rule. In this framework the usual Max Born probability law
\begin{eqnarray}\mathcal{P}^{(\textrm{Born})}_\alpha=\lVert\Psi_\alpha\rVert^2=\lVert\langle\alpha|\Psi\rangle\rVert^2
\end{eqnarray} for observing an outcome $\alpha$ during a quantum measurement seems to conflict with pure unitarity.\\  
\indent Everett~\cite{Everett1957,Barrett2012,DeWitt1973} introduced an additive measure $\mathcal{M}(\lVert\Psi_\alpha\rVert^2)\equiv\lVert\Psi_\alpha\rVert^2$ and subsequently identified it with a probability for the outcome $\alpha$ in the Hilbert space~\footnote{The deduction of Everett must be compared with  the famous Gleason theorem~\cite{Gleason1957,Lubkin1979} which is not valid for an Hilbert space of dimension 2.}. Moreover, identifying the  Everett measure $\mathcal{M}(\lVert\Psi_\alpha\rVert^2)$ with a probability seems \textit{apriori} paradoxical since the MWI doesn't contain chancy events or randomness which could allow us to speak about probability for being this or that. In the  MWI all events occur in parallel and the only certainty is that, after an experiment, we will end-up with a probability $\mathcal{P}=1$ in a superposed quantum state including many branches. In order to solve this contradiction and establish Born's rule Everett based his reasoning on the law of large numbers and the notion of `typicality' advocated by Boltzmann for statistical mechanics \cite{Goldstein2012} (this notion is also assumed in `Bohmian' mechanics \cite{Durr1992}). Indeed, as it is explained in Everett's PhD thesis \cite{Barrett2012} (and as it was more rigorously demonstrated by Hartle \cite{Hartle1968}, DeWitt and Graham~\cite{DeWitt1973,DeWitt1971} and several others\cite{Farhi1989,Aharonov2002}), a long-run experiment reproducing a multinomial sequence leads in the infinite limit, i.e., by a direct application of the law of large numbers, to the empirical statistical Born's rule. In other words, consider \begin{eqnarray}
|\Psi\rangle=\sum_\alpha\Psi_\alpha |\alpha\rangle
\end{eqnarray} a quantum state with outcomes labeled by $\alpha$  which is analyzed in a multi-gates Stern-Gerlach experiment. By taking a long-run sequence of the  same experiment (i.e., by using a tensor product state like $|\Psi_N\rangle\otimes:=|\Psi^{(1)}\rangle\otimes...\otimes|\Psi^{(N)}\rangle$) Everett was able to obtain the relation
\begin{eqnarray}\mathcal{M}(\lVert\Psi_\alpha\rVert^2)=\lim_{N \to +\infty}\frac{N_\alpha}{N}:=\mathcal{P}_\alpha\end{eqnarray} 
i.e., he was able to identify his measure to the relative frequency of occurrence $\frac{N_\alpha}{N}$, where $N_\alpha$ is the number of times the outcome $\alpha$ occurred in the long-run sequence with $N\rightarrow +\infty$ repetitions. Everett here used a standard frequentist definition of probability relying on an infinite ensemble but he also motivated his reasoning with Bayesian and epistemic concepts. In \cite{Barrett2012} he wrote:
\begin{quote}
\textit{We are then led to the novel situation in which the formal theory is objectively continuous and causal, while subjectively discontinuous and probabilistic.}~\cite{Barrett2012}, p.~9. 
\end{quote} 
To understand his reasoning suppose, that $h:=[\alpha_1$,... $\alpha_N]$ is a sequence of outcomes, i.e., an `history'\footnote{ The notion of history used here is reminiscent of GellMann and Hartle work   in the context of the consistent/decoherent histories interpretation \cite{GellMann}. Here, we use history for either a chronological series or for describing a large ensemble  of $N$ identical subsystems at a given time. The results would be the same since the subsystems are factorized and non-interacting.}. For instance, consider an observer (named Alex) participating to the unitary evolution of a quantum measurement with $N$ repetitions. We have: 
\begin{eqnarray}
|\Psi_N\rangle\otimes|\textrm{Alex}_0,E_0\rangle\rightarrow \sum_h|\Psi(h)\rangle \otimes|\textrm{Alex}_h,E_h\rangle
\label{alex}
\end{eqnarray} where Alex$_h$ is Alex$_0$ successor having  a memory of the particular $h-$history in the whole sum and where $|\Psi_N\rangle\otimes=\sum_h|\Psi(h)\rangle $ is the sum of the quantum histories $h$. By subjective probability Everett actually meant  something which is directly measurable by the observer in the branch $h$ where (s)he is located, i.e.,  completely ignoring the existence of the other decohered branches.  In other words, for Everett the natural subjective probability is the limit frequency  $\frac{N_\alpha(h)}{N}$ where $N_\alpha(h)$ is the number of times the event $\alpha$ was repeated in the history $h$. In the $N\rightarrow +\infty$ limit Everett shows  that the total measure $\delta\mathcal{M}$ associated with histories $h$ not fulfilling Born's rule is `overwhelmingly' smaller than the measure associated with the set of histories satisfying Born's rule. At the limit $N\rightarrow +\infty$ the fraction goes to zero \footnote{If we use the sequence $|\Psi_N\rangle=|\Psi^{(1)}\rangle\otimes...\otimes|\Psi^{(N)}\rangle=\otimes_{i=1}^{N}|\Psi^{(i)}\rangle$ with $|\Psi^{(i)}\rangle=\sum_\alpha\Psi_\alpha |\alpha^{(i)}\rangle$ we can define a frequency operator as 
\begin{eqnarray}\hat{Q}_\alpha=\sum_{i=1}^{N}\frac{\hat{\Pi}^{(i)}_\alpha}{N}\label{hartle}\end{eqnarray} with the projectors  $\hat{\Pi}^{(i)}_\alpha=|\alpha^{(i)}\rangle\langle\alpha^{(i)}|$ associated with the eigenvalue $\alpha$ for the $i^{th}$ subsystem. We can expand  the total  state $|\Psi_N\rangle$ as a sum over the different histories $h=[\alpha_1,...,\alpha_N]$, i.e.,  $|\Psi_N\rangle=\sum_h|\Psi(h)\rangle$ with the history quantum state:
\begin{eqnarray}
|\Psi(h)\rangle=\bigotimes_{i=1}^{N}\hat{\Pi}^{(i)}_{\alpha_i}|\Psi_N\rangle=\Pi_{\alpha}\Psi_\alpha^{N_\alpha(h)}\bigotimes_{i=1}^{N}|\alpha_i^{(i)}\rangle \label{truc}
\end{eqnarray} where $N_\alpha(h)$ is the number of times the outcome $\alpha$ occurs for the specific history $h$.  Applying $\hat{Q}_\alpha$ on $|\Psi(h)\rangle$ leads directly to  
\begin{eqnarray}
\hat{Q}_\alpha|\Psi(h)\rangle=\sum_{i=1}^{N}\frac{\delta_{\alpha,\alpha_i}}{N}|\Psi(h)\rangle=\frac{N_\alpha(h)}{N}|\Psi(h)\rangle
\end{eqnarray} where $\sum_{i=1}^{N}\frac{\delta_{\alpha,\alpha_i}}{N}=\frac{N_\alpha(h)}{N}$ appears as an eigenvalue. From the point of view of the observer memory having access to only one of the various  histories $h$ the number $N_\alpha(h)$ is all what is empirically and `subjectively' available. However, for comparing the various histories and objectively define a criterion for the `likelihood'  we still need the measure $\lVert|\Psi(h)\rangle\rVert^2$. For example the average on the whole ensemble leads to $\langle\Psi_N|\hat{Q}_\alpha|\Psi_N\rangle=\sum_h\langle\Psi(h)|\hat{Q}_\alpha|\Psi(h)\rangle=\sum_h\frac{N_\alpha(h)}{N}\lVert|\Psi(h)\rangle\rVert^2 =\lVert\Psi_\alpha\rVert^2$ which is the standard quantum result. This could be made even more precise by reintroducing the notion of typicality in the $N\rightarrow +\infty$ limit. Then a typical history $\bar{h}$ seen  by a typical observer will confirm the record $\frac{N_\alpha(\bar{h})}{N}\simeq\lVert\Psi_\alpha\rVert^2 $ with a error which is going like $\Delta N_\alpha/N_\alpha(\bar{h})\simeq \frac{1}{\sqrt{N}}\sqrt{\frac{1-\lVert\Psi_\alpha\rVert^2}{\lVert\Psi_\alpha\rVert^2}}\rightarrow 0$.\label{foot1}}. Moreover, a typical history $\bar{h}$ seen  by a typical observer entangled with the system confirms the record $\frac{N_\alpha(\bar{h})}{N}\simeq\lVert\Psi_\alpha\rVert^2 $ with a error which is going like $\Delta N_\alpha/N_\alpha(\bar{h})\simeq \frac{1}{\sqrt{N}}\sqrt{\frac{1-\lVert\Psi_\alpha\rVert^2}{\lVert\Psi_\alpha\rVert^2}}\rightarrow 0$. Therefore, the Born rule is `typical' in the Boltzmann sense since the overwhelming majority of the history space (i.e.  weighted with Everett's measure $\mathcal{M}$) is filled by terms satisfying the probability rule of quantum mechanics. This great result has been called the meta-theorem by DeWitt \cite{DeWitt1971} but, as previously explained, it was already discussed by Everett (this notion of `almost all except for a set of measure nearly equals to zero' was considered by Everett as the core of his thesis). The theorem relies critically on the definition of an actually infinite sequence which is never encountered in the lab  and therefore the deduction was often criticized for being circular~\cite{Ballentine1973,Kent1990,Squires1990}. However, this problem is not so fundamental and is actually generic of the application of the law of large numbers in statistical mechanics through the introduction of collectives or Gibbs ensembles (recently some attempts have been made for making sense of such an infinite sequence $N\rightarrow +\infty$ in the MWI, i.e., by linking the problem with the notion of Multiverses used in cosmology~\cite{Aguire2011}).\\
\indent Moreover, the real issue in Everett reasoning concerns the status and unicity of the Everett measure $\mathcal{M}(\lVert\Psi_\alpha\rVert^2)$ for this quantum branching. To paraphrase Wallace~\cite{Wallacevideo} what is only proven by the Everett `law of large numbers' is that relative frequency tends to weight with high weight... Therefore justifying the choice $\mathcal{M}(\lVert\Psi_\alpha\rVert^2)\equiv\lVert\Psi_\alpha\rVert^2$ is central in order to avoid circularity. Yet, it is known in the context of the pilot-wave interpretation (PWI) \cite{deBroglie,BohmHiley},  i.e., in de Broglie Bohm (aka Bohmian) mechanics, that Everett's choice for the measure is far from being univocal (this point was already stressed by Pauli \cite{Pauli} as an objection to Bohm's theory and it becomes the core of recent Valentini's studies~\cite{Valentini} about quantum non-equilibrium in the PWI).  Moreover, changing the measure also changes the notion of typicality and the convergence to a different probability rule \footnote{In Bohmian mechanics we shows~\cite{Valentini} that the density of probability $\rho(X,t)$ in the configuration space reads generally $f(X,t)\lvert\Psi(X,t)\rvert|^2$ where $f(X,t)$ satisfies the relation $\frac{d}{dt}f(X_\Psi(t),t)=0$ along the Bohmian paths $X_\Psi(t)$. This result keeps its importance in the MWI since it shows that any derivation of unicity of Born's rule is necessarily circular.}. This problem also occurs in the context of the MWI and the Everett  weight is clearly not the unique possibility for defining a probability measure. The most natural weight in the context of the MWI would be perhaps the simple branch counting but it is known since Graham \cite{DeWitt1973} that this measure generally conflicts with  Born's rule \footnote{In particular a simple branch counting is not time invariant \cite{Wallace2012,Bricmont2016}. Additionally it requires a preferred basis which must be chosen perhaps in relation with decoherence or the observer memory states.}. Importantly, in the PWI the ontology of the theory concerns (at least in the non-relativistic regime) the particle positions $X_t$ in the configuration space at time $t$. The distribution of particles defines an additional ontological structure absent in the MWI~\cite{DrezetIJQF}.  \\
\indent Moreover, in the recent years the persistent difficulty about defining what probability \emph{actually means} in the MWI has been called the \textit{incoherence problem} and it still plagues any serious discussion about probability in this theory. For instance Albert wrote:
\begin{quote}
\textit{The questions to which this program is addressed are questions of what  we would do if we believed that the fission hypothesis were correct. But the question at issue here is precisely whether to believe that the fission hypothesis is correct!}~\cite{Albert2015} 
\end{quote} In this optics the Everett measure is at best interpreted as an intensity of the ontological state: `a measure of existence' as it is often called by Vaidman~\cite{Vaidman1998}, and the physical interpretation is still contentious after 60 years since Everett's work. Vaidman \cite{Vaidman1998,Vaidman2012,Vaidman2014,Vaidman2020} and Tappenden~\cite{Tappenden2010,Tappenden2019} for instance, propose to introduce Born's rule as an added postulate (under the name \emph{probability postulate} or Born-Vaidman rule~\cite{Tappenden2010,Vaidman2020}) for assigning a degree of \emph{subjective location uncertainty} to the observer in her/his history $h$ (Greaves \cite{Greaves2004} also speaks about caring measure but the meaning is actually a bit different). However, the exact, physical and empirical meaning of the word probability used in this interpretation has been strongly criticized by Albert~\cite{Albert2015}, Kent~\cite{Kent1990} and Maudlin~\cite{Maudlin2019} (see also the discussions in \cite{bookMWI})\footnote{For similar reasons (and many others that we will not review) J.S. Bell \cite{Bell2004}, p.~192 dubbed the MWI a `romantic counterpart of the pilot wave picture' since despite all its glamorous aspects at first sight it can not apparently be developed into a sharp theoretical framework avoiding internal physical contradictions.}. As emphasized by Albert~\cite{Albert2015} this notion of subjective self-locating  uncertainty contrasts and conflicts  with the absence of objective uncertainty in the MWI (indeed everything is unitary so there is nothing of fundamentally uncertain). For Albert, it is only by confusing these two notions of subjective (epistemic) and objective (statistical) probability that one could hope to see a virtual solution where probably there is none. Indeed, the subjective self-locating uncertainty is an internal `pattern'~\cite{Wallace2012} of the observer history $h$. There is thus \emph{a priori} no reason for introducing an objective `caring measure' \cite{Greaves2004} or a `measure of existence' \cite{Vaidman1998,Vaidman2014} in order to weight this subjective notion of uncertainty. However, we stress that the Born-Vaidman axiom can be a posteriori justified by using locality and symmetry for branches with equal weights~\cite{Vaidman2012,Vaidman2020} (i.e., in order to agree with empirical evidences). Furthermore, it  provides an objective structure to the branches of the wave-function and at the same time (i.e., by using the `principal principle' of Lewis~\footnote{The general idea behind a subjectivist approach of probability is to define a degree of belief or credence $\mathcal{C}_\alpha$ for the occurrence of the outcome $\alpha$. Moreover, following  the philosopher D.~Lewis we might use the so called `\emph{principal principle}' for equaling this subjective likelihood to an objective weight  playing also the role of probability $\mathcal{P}_\alpha$. More precisely, the credence assigned to the realization of the outcome $\alpha$ and  conditioned on the knowledge of the objective probability $\mathcal{P}_\alpha$ equals $\mathcal{P}_\alpha$: i.e.,  $\mathcal{C}(\alpha|\mathcal{P}_\alpha)\equiv\mathcal{P}_\alpha$.}) it supplies a subjective probability assignment (or degree of belief)  matching the objective measure. Yet, if the Born-Vaidman rule is a postulate that must be added to the MWI in order to recover standard's quantum mechanics it also implies that unitarity alone is not enough to explain and justify probability.  This motivates the present work  since our proposal   is to supplement  the bare Everett theory, assuming only unitarity, with a (toy) model for quantum observers and minds: a unitary MMI. This model will provide a dynamical structure that in turn legitimize the use of the Born-Vaidman rule. In the next section we will first describe the original stochastics MMI.      
\section{About many-minds} \label{sec2b} 
\indent A very different strategy, which goes back to the late  Zeh in the 1970's~\cite{Zeh} and was subsequently developed by Albert and Loewer in 1988~\cite{AlbertLoewer1988,Albert1994,Barrett1995}, is the the so called MMI which we will shortly describe below. In brief, the idea is to include the role of states of consciousness or awareness into the quantum game. At the difference of older attempts in the same vein such as the von Neumann~\cite{vonNeumann}, London and Bauer~\cite{LondonBauer} and Wigner~\cite{Wigner} approaches, the  MMI involves several mind states $\mathcal{O}^{(1)},\mathcal{O}^{(2)},...$ associated with a single observer. In the approach advocated by Albert and Loewer such mind states are not obeying to the unitary Schr\"{o}dinger equation but are nevertheless guided by solutions $\Psi_t$ of such an equation. Again, in complete analogy with the PWI the mind states associated with the brain structure surf  on the $\Psi_t$ associated with the entangled wave-functions coupling the observer to the measurement apparatus and the quantum object under studies. By surfing on the pilot-wave the many mind states $\mathcal{O}^{(i)}$, which are associated with a given observer and which are unaware from  each other, are stochastically driven into the distinct grooves and channels associated with the wave-function branches. If the wave-function for the observed system reads as before $|\Psi\rangle=\sum_\alpha\Psi_\alpha |\alpha\rangle$ the MMI of Albert and Loewer postulates that a fraction  $\mathcal{P}_\alpha=\lVert\Psi_\alpha\rVert^2$ of mind states given by Born's rule is stochastically driven in the groove, i.e., world corresponding to the outcome $\alpha$. Consider for example a simple non-symmetric  beam splitter experiment  where the quantum state of let say a single photon or electron  evolves as 
\begin{eqnarray}
|\Psi_0\rangle\rightarrow|\Psi_t\rangle=\sqrt{\frac{1}{3}}|\uparrow\rangle+\sqrt{\frac{2}{3}}|\downarrow\rangle\label{Albertstate0} 
\end{eqnarray} where $\uparrow$ and $\downarrow$ describes the two states of the single particle transmitted or reflected by the beam splitter.
 In a more realistic way of describing the experiment we must include an observer (Alex), and an experimental environment (E) into the unitary evolution reading now: 
\begin{eqnarray}
|\Psi_0\rangle\otimes |E_0,\textrm{Alex}_0\rangle\rightarrow\sqrt{\frac{1}{3}}|\uparrow\rangle\otimes |E_\uparrow,\textrm{Alex}_\uparrow\rangle+\sqrt{\frac{2}{3}}|\downarrow\rangle\otimes |E_\downarrow,\textrm{Alex}_\downarrow\rangle.\label{Albertstate} 
\end{eqnarray}  Here, the observer has a memory or record of the experimental outcome as indicated by the $\uparrow\downarrow$ label. In the MMI proposed by Albert and Loewer we add mind states   moving stochastically. For example with one single mind state we have either
 \begin{eqnarray}
|\Psi_0\rangle\otimes |E_0,\textrm{Alex}_0(\mathcal{O}_0^{(1)})\rangle\rightarrow\sqrt{\frac{1}{3}}|\uparrow\rangle\otimes |E_\uparrow,\textrm{Alex}_\uparrow(\mathcal{O}_\uparrow^{(1)})\rangle \nonumber\\ +\sqrt{\frac{2}{3}}|\downarrow\rangle\otimes |E_\downarrow,\textrm{Alex}_\downarrow\rangle \label{first}
\end{eqnarray}  if the mind state moves randomly to the brain of Alex seeing the $\uparrow$-photon or alternatively
\begin{eqnarray}
|\Psi_0\rangle\otimes |E_0,\textrm{Alex}_0(\mathcal{O}_0^{(1)})\rangle\rightarrow\sqrt{\frac{1}{3}}|E_\uparrow,\uparrow\rangle\otimes |\textrm{Alex}_\uparrow\rangle \nonumber\\ +\sqrt{\frac{2}{3}}|\downarrow\rangle\otimes |E_\downarrow,\textrm{Alex}_\downarrow(\mathcal{O}_\downarrow^{(1)})\rangle \label{second}
\end{eqnarray} if the mind state moves along the second groove or branch of the wave-function. The probability of the first alternative (i.e., Eq.~\ref{first}) is $\mathcal{P}_\uparrow=\frac{1}{3}$  whereas for the second alternative (i.e., Eq.~\ref{second}) we have  $\mathcal{P}_\downarrow=\frac{2}{3}$.  By repeating many times the same experiment (and admitting the information about the results of previous experiments are saved) Alex experimentally obtain the Born law by a direct application of the law of large numbers. More precisely, considering a Bernoulli sequence where we repeat $M$ times the previous experiment  we define the probability for having  $M_\uparrow$ times the outcome $\uparrow$ (similarly $M_\downarrow=M-M_\uparrow$) by the binomial formula:
\begin{eqnarray}
\mathcal{P}(M_\uparrow,M_\downarrow)=\frac{M!}{M_\uparrow !M_\downarrow !}\mathcal{P}_\uparrow^{M_\uparrow}\mathcal{P}_\downarrow^{M_\downarrow}\label{retrucmuca}.
\end{eqnarray} where $\mathcal{P}_\uparrow^{M_\uparrow}\mathcal{P}_\downarrow^{M_\downarrow}$ is the probability of an alternative.  By maximizing $\mathcal{P}(M_\uparrow,M_\downarrow)$ in the $M\rightarrow+\infty $ limit we deduce
\begin{eqnarray}
\mathcal{P}_\uparrow\simeq \frac{\tilde{M}_\uparrow}{M}, \mathcal{P}_\downarrow\simeq \frac{\tilde{M}_\downarrow}{M}\label{retrucmucaproba}
\end{eqnarray} where $\tilde{M}_\uparrow$ and $\tilde{M}_\downarrow$ are numbers  for typical branches where Born's rule holds.\\ 
\indent Importantly, in this theory there is no supervenience of the mental states on the brain and more generally all branches but one of the quantum evolution tree contain no mind state (even though the observer brain exists in all these mindless branches). In order to avoid having endless discussions about various issues raised by `mindless-Hulks' (e.g., if a mindless Alex state is discussing with a second observer) Albert and Loewer suggested the introduction of several mind states existing in parallel in the observer brain and also moving randomly. For example with two mind states $\mathcal{O}^{(1)}$ and  $\mathcal{O}^{(2)}$ the initial quantum state reads  $|\Psi_0\rangle\otimes |E_0,\textrm{Alex}_0(\mathcal{O}_0^{(1)},\mathcal{O}_0^{(2)})\rangle$ and it evolves in one of the four following alternatives:
\begin{eqnarray}
\sqrt{\frac{1}{3}}|\uparrow\rangle\otimes |E_\uparrow,\textrm{Alex}_\uparrow(\mathcal{O}_\uparrow^{(1)},\mathcal{O}_\uparrow^{(2)})\rangle+\sqrt{\frac{2}{3}}|\downarrow\rangle\otimes |E_\downarrow,\textrm{Alex}_\downarrow\rangle, \nonumber \\
\sqrt{\frac{1}{3}}|\uparrow\rangle\otimes |E_\uparrow,\textrm{Alex}_\uparrow(\mathcal{O}_\uparrow^{(1)})\rangle+\sqrt{\frac{2}{3}}|\downarrow\rangle\otimes |E_\downarrow,\textrm{Alex}_\downarrow(\mathcal{O}_\downarrow^{(2)}) \rangle,\nonumber \\
\sqrt{\frac{1}{3}}|\uparrow\rangle\otimes |E_\uparrow,\textrm{Alex}_\uparrow(\mathcal{O}_\uparrow^{(2)}) \rangle+\sqrt{\frac{2}{3}}|\downarrow\rangle\otimes |E_\downarrow,\textrm{Alex}_\downarrow(\mathcal{O}_\downarrow^{(1)})\rangle,\nonumber \\
\sqrt{\frac{1}{3}}|\uparrow\rangle\otimes |E_\uparrow,\textrm{Alex}_\uparrow \rangle+\sqrt{\frac{2}{3}}|\downarrow\rangle\otimes |E_\downarrow,\textrm{Alex}_\downarrow(\mathcal{O}_\downarrow^{(1)},\mathcal{O}_\downarrow^{(2)})\rangle.\label{trucmuc} 
\end{eqnarray}  with probabilities respectively equal to $\mathcal{P}_\uparrow^2=\frac{1}{9}$ for the first alternative,  $\mathcal{P}_\uparrow\mathcal{P}_\downarrow=\frac{2}{9}$ for the second and third alternatives, and $\mathcal{P}_\downarrow^2=\frac{4}{9}$ for the last one.  It is clear that if we have $N$ mind states $\mathcal{O}^{(1)},...,\mathcal{O}^{(N)}$ we have now $2^N$ combinations. The total probability for having  $N_\uparrow$ mind states in the upper branch and $N_\downarrow=N-N_\uparrow$ mind states in the lower branch is given (again) by the binomial formula:
\begin{eqnarray}
\mathcal{P}(N_\uparrow,N_\downarrow)=\frac{N!}{N_\uparrow !N_\downarrow !}\mathcal{P}_\uparrow^{N_\uparrow}\mathcal{P}_\downarrow^{N_\downarrow}\label{retrucmuc}.
\end{eqnarray} where $\mathcal{P}_\uparrow^{N_\uparrow}\mathcal{P}_\downarrow^{N_\downarrow}$ is the probability of an alternative.  By maximizing $\mathcal{P}(N_\uparrow,N_\downarrow)$ in the $N\rightarrow+\infty $ limit we obtain  
\begin{eqnarray}
\mathcal{P}_\uparrow\simeq \frac{\tilde{N}_\uparrow}{N}, \mathcal{P}_\downarrow\simeq \frac{\tilde{N}_\downarrow}{N}\label{retrucmubis}
\end{eqnarray} where $\tilde{N}_\uparrow$ and $\tilde{N}_\downarrow$ are the typical numbers of mind states in the upper and lower branch respectively. Therefore, the minds are distributed according to Born's rule. Furthermore, the relative fluctuation to this optimum (written $\frac{\Delta N_\uparrow}{\tilde{N}_\uparrow}=\frac{1}{\sqrt{N}}\sqrt{\frac{\mathcal{P}_\downarrow}{\mathcal{P}_\uparrow}}$)  goes to  zero as $N\rightarrow+\infty $ and the probability to have maverick branches without mind state tends to vanish. This implies that a second observer (Boris) discussing with Alex about his experimental results will always be in contact with $\tilde{N}_\uparrow \gg 1$ Alex minds if the result $\uparrow$ ocurred in his branch (or similarly $\tilde{N}_\uparrow \gg 1$ Alex minds if the result $\uparrow$ ocurred).  Therefore, Boris will typically never meet a mindless hulk (as required). Formally, it means that the probability for the $j^{th}$ mind of Boris recording $\uparrow$ to meet $\tilde{N}_\uparrow=N\mathcal{P}_\uparrow$ Alexs is $\mathcal{P}(\mathcal{O}_\uparrow^{(Boris,j)} \textrm{ to meet } \tilde{N}_\uparrow \mathcal{O}_\uparrow^{(Alex)})\simeq 1$.\\
 \indent Moreover, consider a pair of observers Alex and Boris for the same experiment.  If one mind of Boris during a long Bernoulli sequence records an history like $h=[\uparrow,\downarrow...]$ (where, in agreement with Eq. \ref{retrucmucaproba}, we have $\mathcal{P}_\uparrow\simeq \frac{\tilde{M}_\uparrow}{M}$, 
$\mathcal{P}_\downarrow\simeq \frac{\tilde{M}_\downarrow}{M}$) thus we expect at least one mind in Alex brain recording the same sequence $h$ (since otherwhise Bob and Alex would have no mutual agreement about the result).  The constraints for this correspondence to happen are however very stringent. Indeed, now we have to consider an ensemble of $M$ repetitions for $N$ Alex-minds and the probability for each of these minds to have an history like $h$ reads $\mathcal{P}_h=\mathcal{P}_\uparrow^{\tilde{M}_\uparrow}\mathcal{P}_\downarrow^{\tilde{M}_\downarrow}$. Across the $N$ minds we now get a multinomial probability law   \begin{eqnarray}
\mathcal{P}(\{N_h\})=\frac{N!}{\Pi_h N_h !}\Pi_h\mathcal{P}_h^{N_h}
\end{eqnarray} where $N_h$ is the number of minds having observed the sequence $h$. The law of large numbers give you the result
\begin{eqnarray}
\mathcal{P}_h\simeq \frac{\tilde{N}_h}{N}
\end{eqnarray}which is in general a very small number. For instance, if $\mathcal{P}_\downarrow=1/2$ we have $\mathcal{P}_h=1/2^M$ and therefore we get $\tilde{N}_h=\frac{N}{2^M}$. If we want   $\tilde{N}_h$ finite then at least  $N\sim 2^M$.  Now, this requires a gigantic number of minds since  $N$ can be quite large! This is an issue that a good MMI should clarify one day if the model wants to be taken seriously.        \\
\indent The previous example can easily be generalized to a quantum state like $|\Psi_t\rangle=\sum_\alpha\Psi_\alpha |\alpha\rangle$ which (together with the environment+observer) evolves as 
\begin{eqnarray}
|\Psi_0\rangle\otimes |E_0,\textrm{Alex}_0\rangle\rightarrow \sum_\alpha\Psi_\alpha |\alpha\rangle\otimes |E_\alpha,\textrm{Alex}_\alpha \rangle.
\end{eqnarray}  After the inclusion of $N$ observer mind states and after redoing the previous reasoning we have a multinomial probability for having the set of $\{N_\alpha\}$ mind states (i.e., with $N=\sum_\alpha N_\alpha$): 
  \begin{eqnarray}
\mathcal{P}(\{N_\alpha\})=\frac{N!}{\Pi_\alpha N_\alpha !}\Pi_\alpha\mathcal{P}_\alpha^{N_\alpha}
\end{eqnarray} which is leading again to  the relative frequency of mind states $\mathcal{P}_\alpha:=\lVert\Psi_\alpha\rVert^2 \simeq \frac{\tilde{N}_\alpha}{N}$ for the typical configuration in agreement with Born's rule (the fluctuation reads now $\frac{\Delta N_\alpha}{\tilde{N}_\alpha}=\frac{1}{\sqrt{N}}\sqrt{\frac{1-\mathcal{P}_\alpha}{\mathcal{P}_\alpha}}$ that is also vanishing in the $N\rightarrow +\infty$ limit). The MMI by keeping the psycho-physical parallelism at the statistical average level (i.e., as a very good approximation in the $N\rightarrow +\infty$ limit) is thus remarkably able to reproduce standard quantum mechanical results. \\
\indent However, even if the introduction of the observer memory and mind state has a old and respectable tradition in quantum mechanics interpretation (it was also playing a role in the work of Everett himself) it is fair to say that such an odd approach has been watched with suspicion by many in part because the theory is dualistic in spirit  (separating minds from the rest of the unitary evolution in the Universe), i.e., it breaks down the functionalist or psycho-physical parallelism which is generally accepted (see e.g. von Neumann \cite{vonNeumann}) for discussing quantum mechanics of the observer. In other words in the MMI the mind states don't superverne on the brain state even though the psychophysical parallelism holds at the statistical level as explained before. Furthermore, a multiplicity of minds is required for solving the `mindless-hulk' problem at the price of introducing a form of schizophrenic many-worlds.  It has been attempted to eliminate this unwarranted feature by reinstating psycho-physical realism at the mind level (see Lockwood~\cite{Lockwood1989,Lockwood1996a} and Donald~\cite{Donald1,Donald2,Donald3}). In other words, it has been proposed to re-establish the supervenience of the mind state $\mathcal{O}$ on the brain state described quantum mechanically.  The main difficulty with this new amendment of the MWI (see the interesting discussion following \cite{Lockwood1996a}: \cite{Brown1996,Butterfield1996,Deutsch1996,Loewer1996,Saunders1996,Papineau1996,Lockwood1996b}, see also \cite{Papineau}) is that the mind now becomes a deterministic function of the wave-function state $\Psi_t$, i.e., $\mathcal{O}(\Psi_t)$ (instead of having $\Psi_t(\{\mathcal{O}^{(i)}\})$ in agreement with  the theory of Albert and Loewer~\cite{AlbertLoewer1988,Loewer1996}). While this would seem a natural property in a quantum Universe the proponents of the MMI and MWI following this path have not yet been able to justify the Born's rule unambiguously. In Section \ref{sec3} we will develop a  unitary version of the MMI which is free from these contradictions.
\section{Subjective versus objective probabilities: the quantitative problem}\label{sec2d}
\indent  Proponents of the MWI in general split the whole discussion concerning probability into two issues. The first one, that we called the incoherence issue is tied to the mere existence of probability. The second the \emph{quantitative problem} is connected with the specific mathematical and physical justification of the Born rule in the MWI.  Notwithstanding that the incoherence problem has been solved the quantitative problem is fundamentally interesting by itself and motivated most researches in the last decades.\\ 
\indent Here, we discuss the issue by giving a brief introduction to the remarkable works of Deutsch~\cite{Deutsch1999}, Wallace~\cite{Wallace2012} and Zurek \cite{Zurek2005} and to more recent works by Carroll, Sebens~\cite{Sebens2016} and  Vaidman~\cite{McQueen}. D. Deutsch in his seminal article started with decision-theory and attempted to derive the Born rule from non probabilistic axioms of quantum mechanics. As he wrote: 
\begin{quote}
\textit{Thus we see that quantum theory permits what philosophy would hitherto have regarded as a formal impossibility, akin to `deriving and ought to from an is', namely deriving a probability statement from a factual statement. This could be called deriving a `tends to' from a `does'.}~\cite{Deutsch1999} 
\end{quote}    
Clearly this is a very strong claim which is touching both sides of the difficulty, i.e., the incoherence and quantitative issues. Deutsch's proof has been attacked on philosophical and mathematical grounds (see for example \cite{Barnum,Pitowsky}). The incoherence problem will not be further commented~\footnote{The claim has a long tradition in the MWI community. DeWitt for example famously wrote `The mathematical formalism of the quantum theory is capable of yielding its own interpretation'~\cite{DeWitt1971}.}. The formal part of the proof used the notion of Value function  $\mathcal{V}_\Psi $ and utility assigned to a quantum `game', i.e., a quantum experiment. The semantics of classical decision-theory leads to the definition $\mathcal{V}_\Psi(\{x_\alpha \})=\sum_\alpha x_\alpha \mathcal{P}_\alpha$   
where $x_\alpha$ are eigenvalues of the Hermitian operator $\hat{X}^{(S)}=\sum_\alpha x_\alpha \hat{\Pi}^{(S)}_{\alpha}$ acting on the quantum state $|\Psi^{(S)}\rangle=\sum_\alpha\Psi_\alpha |\alpha^{(S)}\rangle$ associated with system S. Assuming a set of decision-theoretic axioms which are non intrinsically probabilistic Deutsch built the probability function $\mathcal{P}_\alpha:=\lVert\Psi_\alpha\rVert^2$ which is identical to Born's rule. The set of axioms was criticized in particular by Barnum et al. \cite{Barnum} who emphasized the existence of an additional permutation symmetry in the derivation (this is strongly connected with the role of entanglement between S and the observer as we show below). This prompted further important works by Wallace and Saunders~\cite{Wallace2003a,Wallace2003b,Wallace2007,Saunders2005,Saunders2008} (see also~\cite{Greaves2004} and \cite{bookMWI} p.~181 and p.~227) who progressively clarified the whole analysis. It leads Wallace to his simple elegant proof \cite{Wallace2012}  which expurgates the reasoning of unwarranted technical sophistications present in the original derivations. Remarkably, in the mean time Zurek \cite{Zurek2003a,Zurek2003b,Zurek2005,Zurek2014} proposed an alternative proof of Born's rule based on \emph{envariance} a neologism for environment-assisted invariance  a purely quantum symmetry based on entanglement of a system with its environment. What is however key here is that Wallace and Zurek proofs are actually isomorphic to one another. I am going to resume briefly Zurek's proof which is capital for my own deduction and then go to Wallace's semantics.  \\ 
\indent Zurek starts with a Schmidt symmetric quantum state 
\begin{eqnarray}
|\Psi^{(SE)}\rangle=\sqrt{\frac{1}{N}}\sum_{\alpha\in\Delta}|\alpha^{(S)}\rangle\otimes|\varepsilon_\alpha^{(E)}\rangle \label{Zurek}
\end{eqnarray} where $S$ denotes the system and $E$ its environment (the basis vectors are orthogonal). The label of the $\alpha$-mode belongs to a set $\Delta$ with cardinality $N$. Zurek introduces swapping operators acting locally on S and reading $\hat{U} ^{(S)}(\alpha\leftrightarrow \beta)=|\alpha^{(S)}\rangle\langle\beta^{(S)}|+H.c.+\hat{R}^{(S)}$ (with $\hat{R}^{(S)}=\hat{I}^{(S)}-|\alpha^{(S)}\rangle\langle\alpha^{(S)}|-|\beta^{(S)}\rangle\langle\beta^{(S)}|$).  We introduce similar operators  $\hat{U} ^{(E)}(\alpha\leftrightarrow \beta)$ for the environment.  Now, as emphasized in \cite{Zurek2003a} applying successively a swap on S and a counterswap on E lets the state invariant, i.e., 
\begin{eqnarray}
\hat{U} ^{(E)}(\alpha\leftrightarrow \beta)\hat{U} ^{(S)}(\alpha\leftrightarrow \beta)|\Psi^{(SE)}\rangle=|\Psi^{(SE)}\rangle.\label{Zurek2}
\end{eqnarray} 
It is a matter of fact (e.g., from the no-signalling theorem~\cite{Barnum2}) that a local action on S should have no-effect on E and therefore assigning \emph{a priori} probability $\mathcal{P}_\Psi(\alpha^{(S)},\varepsilon_\alpha^{(E)})$ to the branch $|\alpha^{(S)}\rangle\otimes|\varepsilon_\alpha^{(E)}\rangle$ in Eq. \ref{Zurek} we must have  after application of $\hat{U} ^{(S)}(\alpha\leftrightarrow \beta)$ on $|\Psi^{(SE)}\rangle$ and by an application of Laplace's principle of indifference the symmetry relation: 
\begin{eqnarray}
\mathcal{P}_\Psi(\alpha^{(S)},\varepsilon_\alpha^{(E)})=\mathcal{P}_{\hat{U}^{(S)}\Psi}(\beta^{(S)},\varepsilon_\alpha^{(E)})\nonumber \\
\mathcal{P}_\Psi(\beta^{(S)},\varepsilon_\beta^{(E)})=\mathcal{P}_{\hat{U}^{(S)}\Psi}(\alpha^{(S)},\varepsilon_\beta^{(E)}).\label{Zurek3}
\end{eqnarray} Here, we have the strong correlations  $\mathcal{P}_\Psi(\varepsilon_\alpha^{(E)}|\alpha^{(S)})=1$, $\mathcal{P}_{\hat{U}^{(S)}\Psi}(\varepsilon_\alpha^{(E)}|\beta^{(S)})=1$ and thus Eq. \ref{Zurek3} actually reads 
\begin{eqnarray}
\mathcal{P}_\Psi(\varepsilon_\alpha^{(E)})=\mathcal{P}_{\hat{U}^{(S)}\Psi}(\varepsilon_\alpha^{(E)})
\nonumber \\
\mathcal{P}_\Psi(\varepsilon_\beta^{(E)})=\mathcal{P}_{\hat{U}^{(S)}\Psi}(\varepsilon_\beta^{(E)})
\label{Zurek3b}
\end{eqnarray}
that is a statement of Laplacian indifference about what is occurring at S for the subsystem E .
By the same token a subsequent application of $\hat{U} ^{(E)}(\alpha\leftrightarrow \beta)$ yields 
\begin{eqnarray}
\mathcal{P}_{\hat{U}^{(S)}\Psi}(\beta^{(S)},\varepsilon_\alpha^{(E)})=\mathcal{P}_{\hat{U}^{(E)}\hat{U}^{(S)}\Psi}(\beta^{(S)},\varepsilon_\beta^{(E)})\nonumber \\
\mathcal{P}_{\hat{U}^{(S)}\Psi}(\alpha^{(S)},\varepsilon_\beta^{(E)})
=\mathcal{P}_{\hat{U}^{(E)}\hat{U}^{(S)}\Psi}(\alpha^{(S)},\varepsilon_\alpha^{(E)}).\label{Zurek4}
\end{eqnarray} The basis of the reasoning is that in Eq. \ref{Zurek3} an hypothetical observer attached to E is indifferent to what is occurring at S (i.e., a swap) whereas in Eq. \ref{Zurek4} an hypothetical observer attached to S is indifferent about the counterswap acting on E \cite{Zurek2003a}. Moreover, this indifference is both subjective (degree of belief $\mathcal{C}$) and objective (physical probability $\mathcal{P}\equiv \mathcal{C}$) and defined by some properties of the system (in agreement with the principal principle). Therefore, here rational agents should conform their credence to physical probabilities. The objectivity is here linked to the Schmidt form of the state which makes the phases of the different branches locally inoperative to S or E (i.e., we have locally a `mixture'). This would not occur without entanglement because interference between branches are in principle possible so that a swap would break the symmetry \cite{Zurek2003a}.  
Regrouping Eqs. \ref{Zurek3} and \ref{Zurek4} and using the fundamental global envariance Eq. \ref{Zurek2} imply directly
\begin{eqnarray}
\mathcal{P}_\Psi(\alpha^{(S)},\varepsilon_\alpha^{(E)})=\mathcal{P}_{\Psi}(\beta^{(S)},\varepsilon_\beta^{(E)}).
\label{Zurek5}
\end{eqnarray} Moreover, the pair of modes $\alpha$ and $\beta$ was arbitrary in the set $\Delta$ and consequently by generalizing to every pairs we deduce the equiprobability condition reading $\mathcal{P}_\Psi(\alpha^{(S)},\varepsilon_\alpha^{(E)})=Const.$. Finally, by normalization we have Born's rule for this special state $|\Psi^{(SE)}\rangle$, i.e., 
\begin{eqnarray}
\mathcal{P}_\Psi(\alpha^{(S)},\varepsilon_\alpha^{(E)})=\frac{1}{N}=\lVert\langle\alpha^{(S)},\varepsilon_\alpha^{(E)}|\Psi^{(SE)}\rangle\rVert^2.
\label{Zurek6}
\end{eqnarray}What is remarkable about this reasoning is its simplicity relying only on quantum symmetries. As stated by Zurek envariance results `from coexistence between perfect knowledge of the whole and complete ignorance of the parts' \cite{Zurek2003a}. Contrarily to classical Laplace's indifference based on ignorance about information which could be in principle recorded and recovered here the indifference is more fundamental and linked to the entanglement of the system~\cite{Zurek2005}. Indeed, there is no hidden-variable in this approach and nothing more fundamental to find out that the quantum symmetry of the system under swap and counterswap which are local operations acting on S or E. \\ 
\indent This point was also emphasized by Wallace who explained that there is perfect symmetry between the outcomes and thus that the ignorance considered here must be genuinely quantum~\cite{Wallace2012}. A quantum gambler (observer) attached to S or E acting on her/his own subsystem will bet rationally on the different outcomes by using Laplace's indifference as explained previously. Therefore, the decision-theoretic scenario proposed by Wallace and Deutsch reduces to the one made by Zurek (it is interesting to point out that the hidden symmetry contained in Deutsch's proof and which was discovered by Barnum et al. \cite{Barnum} is precisely envariance). For the seek of clarity we postpone to  the end of this section a `derivation' of Wallace's proof using the semantics and logic of Zurek formalism.\\ 
\indent We stress that the previous analysis  assumed some basic physical properties which could at first look innocuous but are actually playing a crucial role.  Indeed, the indifference postulate is motivated by three `bare facts' concerning unitarity and locality i.e., facts about the irrelevance of swap on a subsystem S (or E)  on the physical properties of the subsystem E (or S) (for a discussion of how the conjunction of Zurek's facts lead to Born's rule see~\cite{Zurek2005}).  This is reminiscent either from non-signalling (as already briefly alluded and discussed in \cite{Barnum2}) or from a `natural' postulate concerning `knowledge about the whole versus ignorance of the parts' \cite{Zurek2003a}.  Actually, this axiom hides the notion of mixture and reduced density matrix which already assumes the notion of probability to be derived (i.e., the incoherence problem still holds in this formulation). However, this issue is not so harmful if we consider only the quantitative problem independently of the incoherence one, i.e.,  if we are only interested in recovering Born's rule assuming that probabilities already exist. From this perspective Zurek's axiom only told that beyond assuming the mere existence of probability one must additionally postulate the `strong'\footnote{Such conditions are clearly stronger that mere no-signalling which only requires
$\langle \hat{\mathcal{O} }^{(E)}\rangle_\Psi=\langle \hat{\mathcal{O} }^{(E)}\rangle_{\hat{U}^{(E)}\Psi}
$
where $\hat{\mathcal{O} }^{(E)}$ is any local Hermitian operator acting on E solely and $\hat{U}^{(E)}$ is any unitary transformation acting on the environement E (a similar equation with the role of E and S reverted also holds true). }  symmetry which from standard mechanics reads: 
\begin{eqnarray}
\langle \hat{\Pi }_\alpha^{(S)}\otimes\hat{\Pi}_{\varepsilon_\alpha}^{(E)}\rangle_\Psi=\langle \hat{\Pi }_{\hat{U}^{(S)}\alpha}^{(S)}\otimes\hat{\Pi }_{\varepsilon_\alpha}^{(E)}\rangle_{\hat{U}^{(S)}\Psi}\nonumber \\
\label{Zurek0}
\end{eqnarray}
where $\hat{\Pi }_{\hat{U}^{(S)}\alpha}^{(S)}$ is a `causal' notation for the projector  $\hat{U}^{(S)}|\alpha^{(S)}\rangle\langle\alpha^{(S)}|\left .\hat{U}^{(S)}\right.^\dagger=|\beta^{(S)}\rangle\langle\beta^{(S)}|$. This just leads to Eq. \ref{Zurek3} and similar expressions could be used to  obtain Eq. \ref{Zurek4}. These strong symmetries naturally allow us to recover to equiprobability which is indeed the reasoning of Zurek. Therefore, while in the orthodox interpretation  Eq. \ref{Zurek0} follows from the symmetries of the Schmidt quantum state Eq. \ref{Zurek} together with the already assumed Born's rule (i.e., here equiprobability), in the axiomatic of Zurek it is enough to use Eqs. \ref{Zurek3}, \ref{Zurek4} for recovering equiprobability (i.e., Born's rule) and thus avoiding circularity.\\ 
\indent The previous analysis focused on the simple equiprobable case  where $|\Psi^{(SE)}\rangle$ is given by Eqs. \ref{Zurek}. In order to generalize the deduction to any Schmidt state Zurek used a `trick' \cite{Zurek1998} (see also Deutsch \cite{Deutsch1999}) that consists into applying a fine graining procedure. We start with a S state $|\Psi^{(S)}\rangle=\sum_{a}\sqrt{\mathcal{P}_a}|a^{(S)}\rangle$ where $\mathcal{P}_a=\frac{N_a}{N}$ is a rational number. Entanglement with the environment E leads to 
\begin{eqnarray}
|\Phi^{(SE)}\rangle=\sum_{a}\sqrt{\mathcal{P}_a}|a^{(S)}\rangle
\otimes|e_a^{(E)}\rangle \label{Zurek7}
\end{eqnarray} 
We thus introduce the new vectors   
\begin{eqnarray}
|a^{(S)}\rangle=\frac{1}{\sqrt{N_a}}\sum_{\alpha\in\Delta_a}|\alpha^{(S)}\rangle
\label{Zurek8}
\end{eqnarray} and where the cardinality of $\Delta_a$ equals $N_a$ (we have also $\Delta_a\cap\Delta_b=0$ if $a\neq b$ and $\cup_a\Delta_a=\Delta$ with  $\Delta$ the set of all vectors $|\alpha^{(S)}\rangle$ with cardinality $N$). We have thus 
 \begin{eqnarray}
|\Phi^{(SE)}\rangle=\frac{1}{\sqrt{N}}\sum_{\alpha\in\Delta}|\alpha^{(S)}\rangle
\otimes|e_{a_\alpha}^{(E)}\rangle \label{Zurek9}
\end{eqnarray}  where $a_\alpha=a$ if $\alpha\in\Delta_a$. The last step~\footnote{As remarked by an anonymous referee Zurek's original derivation uses an ancillary system $(C)$ and assumes a new basis in the $(E+C)$ subspace, before performing swaps between $(E)$ and $(C)$ to derive probabilities~\cite{Zurek2003a,Zurek2003b,Zurek2005,Zurek2014}.} consists in a global transformation in the SE system reading $|\alpha^{(S)}\rangle
\otimes|e_{a_\alpha}^{(E)}\rangle \rightarrow |\alpha^{(S)}\rangle
\otimes|\varepsilon_\alpha^{(E)}\rangle $  with $|\varepsilon_\alpha^{(E)}\rangle $ a new environmental basis. We finally obtain 
\begin{eqnarray}
|\Psi^{(SE)}\rangle=\frac{1}{\sqrt{N}}\sum_{\alpha\in\Delta}|\alpha^{(S)}\rangle
\otimes|\varepsilon_\alpha^{(E)}\rangle \label{Zurek10}
\end{eqnarray} which is a Schmidt symmetric state identical to  Eq. \ref{Zurek}. Therefore, Eq. \ref{Zurek6} obtains and we finally get by additivity and application of Laplace's indifference principle:
\begin{eqnarray}
\mathcal{P}_\Phi(a^{(S)},e_a^{(E)})=\sum_{\alpha\in\Delta_a}\mathcal{P}_\Psi(\alpha^{(S)},\varepsilon_\alpha^{(E)})=\frac{N_a}{N}=\lVert\langle a^{(S)},e_a^{(E)}|\Phi^{(SE)}\rangle\rVert^2.
\label{Zurek6}
\end{eqnarray} Continuity establishes the generality of the result for the case where $\mathcal{P}_a$ is a real number \cite{Zurek2003a,Zurek2003b,Zurek2005}.\\ 
\indent The most important part of this proof is the fine-graining procedure which can easily be implemented with beam splitters and unitary gates as shown for example by Vaidman     \cite{McQueen,Vaidman2020}.
However, observe first that this trick requires to have high dimensionality of the Hilbert space for the S subsystem (which is in general true). Second, there is here a form of conspiratorial preparation.  Why indeed should  the distinct beams $|a^{(S)}\rangle$ (which could be located in remote regions of space)  be separated in such a way (i.e., Eq. \ref{Zurek8}) to have equiprobability at the end? Such choice is clearly motivated by the desire to rely on a simple branch counting argumentation to define histories for the observers. Indeed, back to 1985, this solution  naturally avoids the problem existing with the original Deutsch approach \cite{Deutsch1985} (which postulated a density of worlds proportional to $\lVert\Psi_\alpha\rangle\rVert^2$, i.e., different from a naive branch-counting reasoning \footnote{While we can not discuss this issue here we emphasize that the idea of many diverging branches is also advocated in the so called `many-Bohmian-worlds' theory \cite{Tipler2014,Bostrom2014,Sebens2014,Hall2014}.}). In return, the price to be paid  consists in dividing the intensity of the original beams $\mathcal{P}_a$ into many sub-beams of equal intensity $1/N$: a procedure which much be defined in advance by an agent knowing the full properties of the entangled SE system. \\ 
\indent  Zurek's fine-graining trick has been accepted by the proponents of the decision-theoretic approach \cite{Deutsch1999,Wallace2012} as well as by Carroll and Sebens \cite{Sebens2016} which all strongly rely on the methodology offered by Zurek with envariance and also involve the notion of self-location uncertainty (with an interpretation different from the one by Vaidman~\footnote{McQueen and Vaidman \cite{McQueen,Vaidman2020} also rely on the no-signalling theorem in order to prohibit faster-than-light communications. A different interpretation of no-signalling has been previously used by Barnum \cite{Barnum2} to recover Zurek's interpretation. We stress that Vaidman \cite{Vaidman2012,Vaidman2014,McQueen,Vaidman2020} also uses Zurek's fine-graining trick but contrarily to the others doesn't consider entanglement and envariance as important or relevant. The key feature for Vaidman is the symmetry between branches with equal weight and Zurek's fine graining is a natural method for reaching this goal. Additionally, Vaidman doesn't really use the word `proof' in his application of Zurek's trick~\cite{McQueen,Vaidman2020}.}). In particular, Carroll and Sebens \cite{Sebens2016} developed a narrative in which an external observer interacting with a SE system like the one described by $|\Psi^{(SE)}\rangle$ in Eq. \ref{Zurek} is going to assign probabilities to the various outcomes depending on the subsystem S or E (s)he is considering and whether or not the unitary swap $\hat{U}^{(S)}$ or counterswap $\hat{U}^{(E)}$ operations are applied. This interesting narrative (based on a principle named Epistemic Separability Principle or ESP) leads directly through Zurek's and Wallace's argument to the equiprobability condition and then ultimately recovers Born's rule as explained above. We emphasize, that there is a disagreement between Carroll and Sebens \cite{Sebens2016} on the one side, and Kent \cite{Kent2015}, McQueen and Vaidman \cite{McQueen,Vaidman2020} on the other side concerning the role of self-location uncertainty in this analysis (see also \cite{Albert2015}).\\
\indent This finally leads us to Wallace scenario that we analyze using Zurek semantics. Wallace~\cite{Wallace2012} like Deutsch was interested into the observation of the system S by an observer Alex that we can directly identify with the state of the environment E.  Wallace thus considers the following state 
 \begin{eqnarray}
|\Psi^{(SE)}\rangle=\sqrt{\frac{1}{N}}\sum_{\alpha\in\Delta}|\alpha^{(S)}\rangle\otimes|\textrm{Alex}_\alpha^{(E)}\rangle \label{Wallace}
\end{eqnarray} as well as the counterswapped state $\hat{U} ^{(E)}(\alpha\leftrightarrow \beta)|\Psi^{(SE)}\rangle$. If the original state contains the terms 
\begin{eqnarray}
 |\Psi^{(SE)}\rangle=|\alpha^{(S)}\rangle\otimes|\textrm{Alex}_\alpha^{(E)}\rangle+|\beta^{(S)}\rangle\otimes|\textrm{Alex}_\beta^{(E)}\rangle+|R\rangle\label{Wallace1}
\end{eqnarray} 
 the new counterswapped state contains instead the terms
 \begin{eqnarray}
 \hat{U}^{(E)}(\alpha\leftrightarrow \beta)|\Psi^{(SE)}\rangle=|\alpha^{(S)}\rangle\otimes|\textrm{Alex}_\beta^{(E)}\rangle+|\beta^{(S)}\rangle\otimes|\textrm{Alex}_\alpha^{(E)}\rangle+|R\rangle\label{Wallace1b}
\end{eqnarray} where $|R\rangle$ is the `rest' which is irrelevant. Now, by direct application of Laplace's indifference principle we obtain (see Eq.\ref{Zurek3} )
\begin{eqnarray}
\mathcal{P}_\Psi(\alpha^{(S)},\textrm{Alex}_\alpha^{(E)})=\mathcal{P}_{\hat{U}^{(E)}\Psi}(\alpha^{(S)},\textrm{Alex}_\beta^{(E)})\nonumber \\
\mathcal{P}_\Psi(\beta^{(S)},\textrm{Alex}_\beta^{(E)})=\mathcal{P}_{\hat{U}^{(E)}\Psi}(\beta^{(S)},\textrm{Alex}_\alpha^{(E)}).\label{Wallace2}
\end{eqnarray} Unlike Zurek  Wallace didn't used a swap on the S subsystem. Instead, he used a trick by supposing that (i) the subsystem S has also a ground state $|\emptyset^{(S)}\rangle$ and that (ii) we apply the erasing operation  $\hat{U_e}^{(S)}|\alpha^{(S)}\rangle=|\emptyset^{(S)}\rangle$, $\hat{U_e}^{(S)}|\beta^{(S)}\rangle=|\emptyset^{(S)}\rangle$. After application of the erasing  process on the states given by Eqs. \ref{Wallace1}, \ref{Wallace1b} we obtain 
\begin{eqnarray}
 \hat{U_e}^{(S)}|\Psi^{(SE)}\rangle=|\emptyset^{(S)}\rangle\otimes(|\textrm{Alex}_\alpha^{(E)}\rangle+|\textrm{Alex}_\beta^{(E)}\rangle+|R\rangle\nonumber \\ 
 \hat{U_e}^{(S)}\hat{U}^{(E)}(\alpha\leftrightarrow \beta)|\Psi^{(SE)}\rangle=|\emptyset^{(S)}\rangle\otimes(|\textrm{Alex}_\beta^{(E)}\rangle+|\textrm{Alex}_\alpha^{(E)}\rangle+|R\rangle
 \label{Wallace3}
\end{eqnarray} This motivates the set of equations: 
 \begin{eqnarray}
\mathcal{P}_\Psi(\alpha^{(S)},\textrm{Alex}_\alpha^{(E)})=\mathcal{P}_{\hat{U}_e^{(S)}\Psi}(\emptyset^{(S)},\textrm{Alex}_\alpha^{(E)})\nonumber \\ 
\mathcal{P}_\Psi(\beta^{(S)},\textrm{Alex}_\beta^{(E)})=\mathcal{P}_{\hat{U}_e^{(S)}\Psi}(\emptyset^{(S)},\textrm{Alex}_\beta^{(E)})\nonumber \\ 
\mathcal{P}_{\hat{U}^{(E)}\Psi}(\alpha^{(S)},\textrm{Alex}_\beta^{(E)})=\mathcal{P}_{\hat{U}_e^{(S)}\hat{U}^{(E)}\Psi}(\emptyset^{(S)},\textrm{Alex}_\beta^{(E)})\nonumber \\
\mathcal{P}_{\hat{U}^{(E)}\Psi}(\beta^{(S)},\textrm{Alex}_\alpha^{(E)})=\mathcal{P}_{\hat{U}_e^{(S)}\hat{U}^{(E)}\Psi}(\emptyset^{(S)},\textrm{Alex}_\alpha^{(E)}).\label{Wallace4}
\end{eqnarray} Finally, we use the fact that by branch indifference~\cite{Wallace2012} we have 
\begin{eqnarray}
\mathcal{P}_{\hat{U}_e^{(S)}\Psi}(\emptyset^{(S)},\textrm{Alex}_\alpha^{(E)})=\mathcal{P}_{\hat{U}_e^{(S)}\hat{U}^{(E)}\Psi}(\emptyset^{(S)},\textrm{Alex}_\alpha^{(E)}) \nonumber \\
\mathcal{P}_{\hat{U}_e^{(S)}\Psi}(\emptyset^{(S)},\textrm{Alex}_\alpha^{(E)})=\mathcal{P}_{\hat{U}_e^{(S)}\hat{U}^{(E)}\Psi}(\emptyset^{(S)},\textrm{Alex}_\alpha^{(E)})
\label{Wallace5}
\end{eqnarray} to get after combining with Eqs. \ref{Wallace2}, \ref{Wallace4} the result
\begin{eqnarray}
\mathcal{P}_\Psi(\alpha^{(S)},\textrm{Alex}_\alpha^{(E)})=\mathcal{P}_\Psi(\beta^{(S)},\textrm{Alex}_\beta^{(E)})\label{Wallace6}
\end{eqnarray} which is Zurek Eq.~\ref{Zurek5}. By proceeding as in Zurek's case we again obtain equiprobability and thus Born's rule Eq. \ref{Zurek6}.\\
\indent Moreover, to conclude this section it is key to understand that the different strategies made by Zurek, Deutsch or Wallace all rely on some subjectivist and epistemic approaches of probabilities.  Strategies of that kind have a old and respectable tradition going back at least to Bernoulli, Laplace or Poisson. In the 20$^{th}$ century they were strongly advocated by  Borel, de Finetti, Ramsey, Keynes (who named the `indifference principle') and many others like Jeffreys, Savage and Lewis. In particular, the principal principle of Lewis `$\mathcal{C}(\alpha|\mathcal{P}_\alpha)\equiv\mathcal{P}_\alpha$' equaling objective probabilities to subjective credences often assumes that we can define objectively `chances' (i.e., in the so-called objective bayesianism where probabilities are assigned on the basis of the maximum entropy principle).  The strategies of Zurek, Deutsch or Wallace also require such axiomatics, but unfortunately, it is not very clear what an objective chance or probability could be in the MWI (this is the source of the incoherence problem).  Therefore, even if  we consider the formal proofs of Zurek, Wallace Deutsch and others as very interesting for the quantitative problem we still believe that they don't unambiguously clarify the incoherence problem.  This is the motivation for the MMI model presented in the next section.                 	
\section{A deterministic and quantum version of the many-minds interpretation}\label{sec3}    
 \indent It is often assumed  by proponents of the MWI that the introduction of probability is not worse (and perhaps not better) than it is in other interpretations of quantum mechanics or even in other  fields of physical science. The claim goes back to Everett \cite{Everett1957} who saw his measure-theoretic deduction as good as the one used in classical statistical physics. More recently, Papineau \cite{Papineau1996,Papineau} and Wallace \cite{Wallace2012} repeated the same claim that probabilities are very obscure concepts and that the MWI is not in a worst position than for example GRW collapse of Bohmian models are for discussing randomness and chances. Wallace \cite{Wallace2003a} and Zurek \cite{Zurek2003a} following Deutsch \cite{Deutsch1999}, went further by claiming that the genuinely quantum Laplacian indifference, i.e., related to envariance and self-locating uncertainty, provides within the MWI framework a even better basis for a clean foundation of probability than in collapse or Bohmian approaches.\\ 
\indent As we saw there are serious reasons to doubt about the validity of such strong claims. First, in collapse interpretations such as GRW or in the Copenhagen interpretation the notion of infinite sequences is not problematic and positing a frequency law like $\mathcal{P}_\alpha\equiv \lim_{N \to +\infty}\frac{N_\alpha}{N}$ (interpreted probabilistically) means that the systems know stochastically, i.e., on the long run how to behave. The stochastic rules in this single-world approach can be axiomatized and the (weak) law of large numbers allows us to connect probabilities and statistics.  Second, the MMI of Albert and Loewer \cite{AlbertLoewer1988} is also based on a stochastic approach to probability and the model is self-consistent even though strongly dualistic. \\ 
\indent The same is true for the PWI where probability arizes from the initial conditions of similar systems typically distributed over a space-like surface. In the PWI, which like classical mechanics is fully deterministic, one must impose an `equivariant' distribution of particles and fields at one time in order to recover Born's rule at any other times. Like for classical statistical mechanics there are persistent debates about the probabilistic foundations of the PWI but these debates are not about incoherence \emph{per se} (which is actually irrelevant in the de Broglie Bohm framework) but are instead focused on the uniqueness of Born's rule and on the status  of quantum equilibrium versus quantum non-equilibrium particle distributions \cite{Durr1992,Valentini}. The problems are very similar to those existing in statistical thermodynamics for justifying  microcanonical and canonical ensembles and for describing the tendency to reach thermal equilibrium. In particular, we emphasize that there exists what could be called  a minimalist PWI advocated by Bell \cite{Bell2004} p.~129 and Goldstein D\"urr and Zangh\`{i } \cite{Durr1992} and where the Boltzmanian notion of  typicality plays a central role for recovering Born's rule.  This approach starts with the same methodology as in Everett's work \cite{Barrett2012}, i.e., by introducing the preferred Everett measure $\mathcal{M}(h)$ assigned to histories $h$ (here defined in the coordinate configuration space for point-like particles or field variables for continuous Bosonic fields). From the weak law of large numbers we deduce, like in the MWI, that Born's rule holds with a near-unit Everett weight in the limit $N\rightarrow +\infty$.\\ 
\indent Moreover, at the difference of the MWI we can precisely and univocally define what we mean by an actual configuration and this even for a finite $N$. Indeed, taking as an example the non-relativistic de Broglie-Bohm dynamics for a system of $N$ electrons we can write the actual density of Bohmian electrons at the spatial point $\mathbf{q}\in\mathbb{R}^{3}$: 
\begin{eqnarray}
\rho_N(\mathbf{q},t)=\frac{1}{N} \sum^{i=N}_{i=1}\delta^{3}(\mathbf{q}^{(i)}_\Psi(t)-\mathbf{q})\simeq\lVert\Psi(\mathbf{q},t)\rVert^2\label{axiom2}
\end{eqnarray} with $\mathbf{q}^{(i)}_\Psi\in\mathbb{R}^{3} $ some `typical' Bohmian paths for the electrons. The second approximate equality means that Born's rule is accurately valid for this `history'\footnote{This relation is equivalent to the frequency relation $\frac{N_\alpha(h)}{N}\simeq\lVert\Psi_\alpha\rVert^2$ defined for some histories $h$ and which is valid even for $N$ finite. Note that the accuracy increases with $N$ since the highly discrete sum of Dirac peaks approaches a continuous fluid with density $\lVert\Psi(\mathbf{q},t)\rVert^2$. A faster convergence is obtained by limiting our analysis to coarse grained probability functions in some elementary but finite spatial cells.} and the Everett measure provides a quantitative figure of merit for that accuracy in the regime $N\gg 1$. We however emphasize that Everett's measure is not the only possibility in the PWI. This is a indeed a measure dependent problem (associated with the choice of the Universe initial conditions) and constitutes the recurrent issue debated by Valentini on the one side \cite{Valentini} and Goldstein, D\"urr and Zangh\`{i } on the other side~\cite{Durr1992}. Therefore, in this framework changing the measure for a non equivariant one $\neq\lVert\Psi(\mathbf{q},t)\rVert^2$  selects a different set of typical histories in which Born's rule will not hold and corresponding to a different choice for the Universe initial conditions. In other words, in the framework of the PWI Wallace's definition of the law of large numbers: `relative frequency tends to weight with high weight' \cite{Wallacevideo} makes physically sense since the weight can be selected in order to agree with a physical quantity $\rho_N(\mathbf{q},t)$ as defined by Eq. \ref{axiom2} and associated with an actual state of the Universe, i.e.,  $\mathbf{q}^{(1)}_\Psi(t),...,\mathbf{q}^{(N)}_\Psi(t)$  and specific initial conditions.  The problem in the MWI is that such a choice is not univocally accepted and therefore (contrarily to the PWI) the notion of probability is not clearly defined.\\ 
\indent We will not pursue  the discussion about the meaning of probability in the PWI  (for a `balanced' review of the general problem see  \cite{Drezet}). Here, we will instead consider Bohmian and classical statistical mechanics as a motivation for new models applied to the MWI and MMI. In the following we will develop a speculative although mathematically precise and unitary toy model for the MMI (for related ideas see  Lockwood~\cite{Lockwood1989,Lockwood1996a} and Donald~\cite{Donald1,Donald2,Donald3}).\\ 
\indent In our approach we first assume that the `all-is-wave' ontology advocated by Everett is true.  Therefore, the quantum state $\Psi_t$ of the Universe has the same physical meaning as in the MWI. However, in order to recover Born's rule and objective probabilities we propose here to mix the theory with ingredients of the MMI.  This assumes a completely unitary version of the mind that we will describe with an idealized model.  We stress that our speculative toy model of quantum mind states (assuming a purely unitary ontology \`a la Everett) is introduced only in order to recover Born's rule. We don't claim that the Universe should necessarily be populated  with such quantum observers  but only that  a clear probabilistic interpretation of measurements is possible  if we accept their existence. For this purpose, we start  with the Albert and Loewer MMI \cite{AlbertLoewer1988} and go back to the example of Eq.~\ref{Albertstate}. However, now we replace Alex by some collective excitation of a memory device which we write $|E_0,\mathcal{O}_0^{(1)}\rangle$ before the interaction. $\mathcal{O}^{(i)}$  (with here $i=1$) is going to play the role of a single mind in the MMI. We consider a symmetric beam splitter and during the measurement operation we postulate the unitary evolution:  
\begin{eqnarray}
|\Psi_0\rangle\otimes |E_0,\mathcal{O}_0^{(1)}\rangle\otimes |\spadesuit^{(1)}\rangle\rightarrow\sqrt{\frac{1}{2}}\left( |\uparrow\rangle\otimes |E_\uparrow,\mathcal{O}_\uparrow^{(1)}\rangle+|\downarrow\rangle\otimes |E_\downarrow,\emptyset^{(1)}\rangle\right) \otimes |\spadesuit^{(1)}\rangle, \nonumber \\
|\Psi_0\rangle\otimes |E_0,\mathcal{O}_0^{(1)}\rangle\otimes |\heartsuit^{(1)}\rangle\rightarrow\sqrt{\frac{1}{2}}\left( |\uparrow\rangle\otimes |E_\uparrow,\emptyset^{(1)}\rangle+|\downarrow\rangle\otimes |E_\downarrow,\mathcal{O}_\downarrow^{(1)}\rangle\right) \otimes |\heartsuit^{(1)}\rangle,
\nonumber \\
\label{Drezetstate} 
\end{eqnarray}  where $|\spadesuit^{(1)}\rangle$ and $|\heartsuit^{(1)}\rangle$  are two orthogonal normalized states of a single qubit taking part to the interaction. The physical meaning of this qubit is to act as an environement for the mind states and creating a form of `distinguishability' between the different alternatives in Eq. \ref{Drezetstate}.   The exponent $i$ labels the qubit `family' which is related to $\mathcal{O}^{(i)}$. Here we start with only one family but we will have to introduce as many families $i=1,...,N$ as we have observer minds $\mathcal{O}^{(i)}$ (see Eqs. \ref{Drezetstaten0}-\ref{Drezetstaten3} below). The $|E_\downarrow,\emptyset^{(1)}\rangle$ is a specific quantum state of the brain where the memory associated with the mind has been lost or destroyed, i.e., this is a kind of ground state. Yet, the brain and environment $E_\downarrow$ could keep some persistent information about the result $\downarrow$ (this is indicated by the state $E_\downarrow$). A similar comment could be done for $|E_\uparrow,\emptyset^{(1)}\rangle$. We emphasize that the presence of the empty states $|E_\downarrow,\emptyset^{(1)}\rangle,|E_\uparrow,\emptyset^{(1)}\rangle$ reinstores the psychophysical parallelism which was lost in the original MMI~\cite{AlbertLoewer1988}. Indeed, here the empty states are imprinted in the wave-function like empty particle modes in quantum electrodynamics. Here (unlike in~\cite{AlbertLoewer1988}) the active mind and empty states are not dualistically separated from the material world. This is clearly an improvement.   \\
\indent Moreover, note that (in agreement with the MWI) everything is deterministic and unitary in this model: depending on the value of the qubit $|\spadesuit^{(1)}\rangle$ or $|\heartsuit^{(1)}\rangle$ the evolution follows one or the other of the two alternatives. Now, comes the trick we could easily introduce some randomness or molecular chaos concerning the state of the qubit and defined at, let say,  the beginning of the Universe.
\begin{figure}[h]
\begin{center}
\includegraphics[width=10cm]{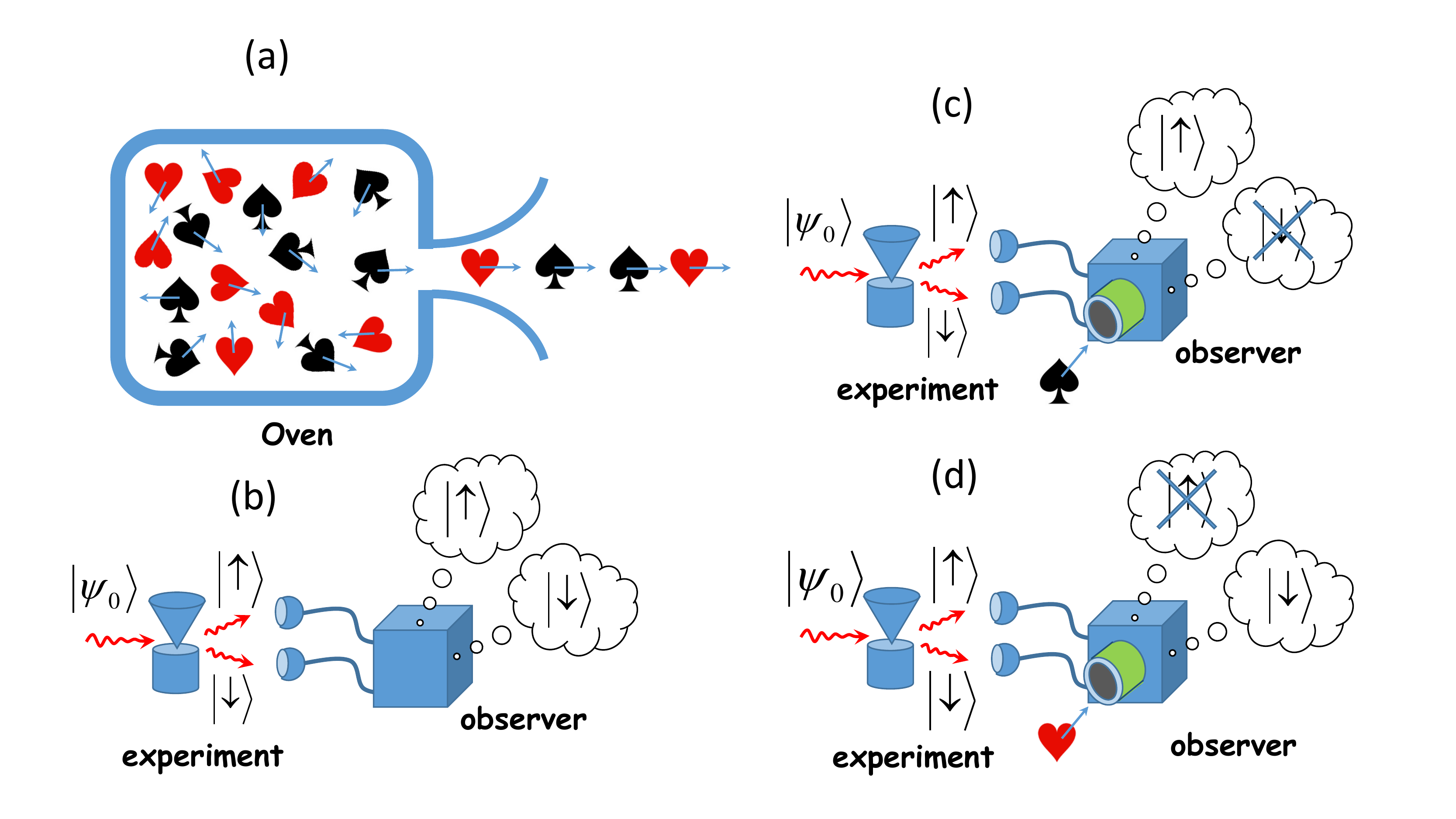} 
\caption{Principle of the quantum single-mind experiment using (a) a random sequence of  $\heartsuit,\spadesuit,\spadesuit,\heartsuit...$ to drive the decision of a quantum machine (`observer') memorizing only one of the two outcomes, i.e., $|\uparrow\rangle$ or $|\downarrow\rangle$ of a quantum experiment (c or d). This situation contrasts with the usual quantum observer, i.e., insensitive to qubits $\heartsuit,\spadesuit,\spadesuit,\heartsuit...$ , which would be in a state of quantum schizophrenia and unable to solve the incoherence problem (b).} \label{figure1}
\end{center}
\end{figure}
More precisely, we assume that we have a large ensemble of $M$ such qubits in a product state like $|h^{(1)}\rangle=|\spadesuit^{(1,1)}\rangle\otimes...\otimes|\heartsuit^{(1,M)}\rangle$ and where the number of $\heartsuit^{(1)}$ typically equals the number of $\spadesuit^{(1)}$, i.e., $\tilde{M}_{\spadesuit^{(1)}}\simeq\tilde{M}_{\heartsuit^{(1)}}$. We have here a `random' but classical distribution of the two states. By doing and redoing the same experiment the observer will interact with one exemplar of the product state (i.e., $\heartsuit^{(1,j)}$ if the j$^{\textrm{th}}$ qubit is a heart or $\spadesuit^{(1,j)}$ if it is a spad).  Therefore, for an observer mind $\mathcal{O}^{(1,j)}$ taken in the ensemble we can objectively define the probability for interacting with a spad or a heart as:   
\begin{eqnarray}
\mathcal{P}_{\spadesuit^{(1)}}:=\frac{\tilde{M}_{\spadesuit^{(1)}}}{M}\simeq \frac{1}{2},&&
\mathcal{P}_{\heartsuit^{(1)}}:=\frac{\tilde{M}_{\heartsuit^{(1)}}}{M}\simeq \frac{1}{2}.\label{thermal}
\end{eqnarray} This probability law could be justified like in classical or Bohmian mechanics by using a typicality approach with equal measures for the two outcomes and by application of the Bernoulli/Laplace law of large numbers in the limit $M\rightarrow +\infty$.\\ 
\indent We emphasize that the typicality reasoning is here classical unlike the one of Everett. Considering all the different possible histories $|h^{(1)}\rangle=\bigotimes_{k=1}^{M}|s_{k}^{(1,k)}\rangle$ (with $s_{k}=$ spade or heart) we need a density matrix~\footnote{Introducing the frequency operator $\hat{Q}_{s^{(1)}}=\sum_{k=1}^{M}\frac{\hat{\Pi}^{(k)}_{s^{(1)}}}{M}$ (see Eq.\ref{hartle} in footnote \ref{foot1}) we have 
$\textrm{Tr}[\hat{Q}_{s^{(1)}}|h^{(1)}\rangle\langle h^{(1)}|]=\frac{M_{s^{(1)}}(h^{(1)})}{M}$  which for typical histories gives us: 
$\frac{M_{s^{(1)}}(\bar{h}^{(1)})}{M}\simeq \frac{1}{2}$. For the whole ensemble we have also $\textrm{Tr}[\hat{Q}_{s^{(1)}}\hat{\rho}^{(1)}]=\frac{1}{2}$.} \begin{eqnarray}
\hat{\rho}^{(1)}=\sum_{h^{(1)}}\mathcal{M}(h^{(1)})|h^{(1)}\rangle\langle h^{(1)}|
\end{eqnarray} with the probability measure $\mathcal{M}(h^{(1)})=\frac{1}{2^M}$. Note, that here probabilities have objective meaning associated with a distribution of particles (i.e., like for a thermal gas), and this approach doesn't require a subjective/epistemic ignorance \`a la Laplace. The actual state of our Universe is one of the typical $|h^{(1)}\rangle$ in the density matrix $\hat{\rho}^{(1)}$. Like for a Gibbs ensemble in statistical mechanics the other alternatives could also actually exist in separated Universes (forming a Gibbs multiverse~\cite{Aguire2011}). Therefore, the introduction of a mixture like $\hat{\rho}^{(1)}$ should not be thought as a breaking of Unitarity but better as a structure added to the MWI in order  to give a classical-like objective meaning to probabilities (i.e., like in the PWI). \\ 
\indent Moreover, from Eq. \ref{Drezetstate} we know that if a qubit is in the state $\spadesuit$ the unitary evolution imposes to the active observer state to be $|E_\uparrow,\mathcal{O}_\uparrow^{(1)}\rangle$ associated with a memory of the $\uparrow$ outcomes. Of course there is also an empty state like  $|E_\downarrow,\emptyset^{(1)}\rangle$ but this not associated with a memory of the observer. We can thus define an objective probability $\mathcal{P}(\mathcal{O}_\uparrow^{(1)})$ that must  corresponds to $\mathcal{P}_{\spadesuit^{(1)}}$. In other words we have:  
\begin{eqnarray}
\mathcal{P}_{\spadesuit^{(1)}}=\mathcal{P}(\mathcal{O}_\uparrow^{(1)})\simeq \frac{1}{2},&&
\mathcal{P}_{\heartsuit^{(1)}}=\mathcal{P}(\mathcal{O}_\downarrow^{(1)})\simeq \frac{1}{2}.
\end{eqnarray}       
  meaning that the probability (i.e., the relative frequency) for the observer mind  $\mathcal{O}^{(1)}$ taken in the ensemble  to memorize the $\uparrow$ (or $\downarrow$) quantum state is one half. This functionalist approach gives also a clear physical meaning to the subjective concept of `degree of belief'. Here, the principal principle identifying objective probabilities and subjective credences is physically justified if we identify the credence $\mathcal{C}(\uparrow|\mathcal{P}(\mathcal{O}_\uparrow^{(1)}))$ with $\mathcal{P}(\mathcal{O}_\uparrow^{(1)})$.     This defines the pivotal result of our deduction for a single mind.\\  
\indent This suggests the following narrative (see Fig. \ref{figure1}).  Imagine a thermal source or oven of (distinguishable) quantum particles with two internal states $\spadesuit$ and $\heartsuit$. The distributions $\frac{\tilde{M}_{\spadesuit}}{M}$ and $\frac{\tilde{M}_{\heartsuit}}{M}$ are given by Eq. \ref{thermal} and justified by using an infinite  Gibbs collective or ensemble like in thermodynamics.   Assuming a hole on the oven  (see Fig. \ref{figure1}(a)) walls we can define a random sequence of escaping particles in the typical ensemble of qubits belonging to the oven (mathematically this corresponds to a fair typical sample taken in a large population).  This typical sequence $h\in [\heartsuit,\spadesuit,\spadesuit,\heartsuit...]$ satisfying Eq. \ref{thermal}  will be used by a quantum observer to decide deterministically if the mind will be aware of the recording of the $|\uparrow\rangle$ or $|\downarrow\rangle$ (see Fig. \ref{figure1} (c,d)). The model with the oven here clearly mirrors the discussion  of the Bernoulli process in the usual MMI of  Albert and Loewer (see the binomial formula Eq.~\ref{retrucmuca}). In the absence of such a mechanism (see Fig \ref{figure1}(b))  the observer mind would be in a schizophrenic superposition of $\uparrow/\downarrow$ information leading to the incoherence difficulty in the MWI. Here we solve the issue for the particular case of a unitary model including minds, i.e., a unitary  MMI.   With the new mechanism proposed here (see Eq. \ref{Drezetstate}) we can objectively give a meaning to randomness in the MMI.  Therefore, here the fundamental quantum randomness of our Universe is revealed in the brain after interaction with qubits $\heartsuit,\spadesuit,\spadesuit,\heartsuit...$ deterministically prepared in a pseudo-random way. This is done without avoiding full unitarity of the quantum evolution (in contrast with the original MMI \cite{AlbertLoewer1988,Albert1994,Barrett1995}). The key feature in this model is the use of molecular chaos associated with the oven.  Ultimately this is associated with a specific choice for the initial conditions of our Universe selecting what is typical or not.  Like in classical and Bohmian statistical mechanics the notion of typicality is clearly connected with such initial conditions.\\     
\indent In a second stage of our analysis we easily extend the pivotal result to an ensemble of many minds.  Consider for example two minds (the generalization being obvious for $N$ minds as we see below). We suppose the initial state  in Eq. \ref{Drezetstate} transformed into  $|\Psi_0\rangle\otimes |E_0,\mathcal{O}_0^{(1)},\mathcal{O}_0^{(2)}\rangle\otimes |s^{(1)}\rangle\otimes|s'^{(2)}\rangle$ where $\mathcal{O}_0^{(1)}$ and $\mathcal{O}_0^{(2)}$  are two mind states (collective excitations) unaware of each other and $|s^{(1)}\rangle\otimes|s'^{(2)}\rangle$ some spin states with $s=$ spade or heart, and $s'=$ spade or heart. After the interaction we obtain four possible outcomes: 
\begin{eqnarray}
|\Psi_0\rangle\otimes |E_0,\mathcal{O}_0^{(1)},\mathcal{O}_0^{(2)}\rangle\otimes |\spadesuit^{(1)}\rangle\otimes |\spadesuit^{(2)}\rangle\rightarrow\nonumber \\
\sqrt{\frac{1}{2}}\left( |\uparrow\rangle\otimes |E_\uparrow,\mathcal{O}_\uparrow^{(1)},\mathcal{O}_\uparrow^{(2)}\rangle+|\downarrow\rangle\otimes |E_\downarrow,\emptyset^{(1)},\emptyset^{(2)}\rangle\right) \otimes |\spadesuit^{(1)}\rangle\otimes |\spadesuit^{(2)}\rangle, 
\\ \label{Drezetstaten0}
|\Psi_0\rangle\otimes |E_0,\mathcal{O}_0^{(1)},\mathcal{O}_0^{(2)}\rangle\otimes |\spadesuit^{(1)}\rangle\otimes |\heartsuit^{(2)}\rangle\rightarrow\nonumber \\
\sqrt{\frac{1}{2}}\left( |\uparrow\rangle\otimes |E_\uparrow,\mathcal{O}_\uparrow^{(1)},\emptyset^{(2)}\rangle+|\downarrow\rangle\otimes |E_\downarrow,\emptyset^{(1)},\mathcal{O}_\downarrow^{(2)}\rangle\right) \otimes |\spadesuit^{(1)}\rangle\otimes |\heartsuit^{(2)}\rangle, 
 \\ \label{Drezetstaten1}
|\Psi_0\rangle\otimes |E_0,\mathcal{O}_0^{(1)},\mathcal{O}_0^{(2)}\rangle\otimes |\heartsuit^{(1)}\rangle\otimes |\spadesuit^{(2)}\rangle\rightarrow\nonumber \\
\sqrt{\frac{1}{2}}\left( |\uparrow\rangle\otimes |E_\uparrow,\emptyset^{(1)},\mathcal{O}_\uparrow^{(2)}\rangle+|\downarrow\rangle\otimes |E_\downarrow,\mathcal{O}_\downarrow^{(1)},\emptyset^{(2)}\rangle\right) \otimes |\heartsuit^{(1)}\rangle\otimes |\spadesuit^{(2)}\rangle,
\\ \label{Drezetstaten2}
|\Psi_0\rangle\otimes |E_0,\mathcal{O}_0^{(1)},\mathcal{O}_0^{(2)}\rangle\otimes |\heartsuit^{(1)}\rangle\otimes |\heartsuit^{(2)}\rangle\rightarrow\nonumber \\
\sqrt{\frac{1}{2}}\left( |\uparrow\rangle\otimes |E_\uparrow,\emptyset^{(1)},\emptyset^{(2)}\rangle+|\downarrow\rangle\otimes |E_\downarrow,\mathcal{O}_\downarrow^{(1)},\mathcal{O}_\downarrow^{(2)}\rangle\right) \otimes |\heartsuit^{(1)}\rangle\otimes |\heartsuit^{(2)}\rangle.  
\label{Drezetstaten3} 
\end{eqnarray}  This situation clearly mirrors the result discussed in Section \ref{sec2b} surrounding Eq. \ref{trucmuc}. Now, like for the single mind  problem we can introduce product states  $\bigotimes_{k=1}^{M}(|s_{k}^{(1,k)}\rangle\otimes|s_{k}^{(2,k)}\rangle)$ with $s_{k}=$ spade or heart. Again we suppose a `typical' random distribution of spades and hearts such that in the product state the population is equally distributed. We have a kind of `\emph{Stosszahlansatz}' (molecular chaos hypothesis) discussed in Boltzmann statistical mechanics and which involves an hypothesis about the absence of correlation (i.e. independence) between the $|s_{k}^{(1,k)}\rangle$ and $|s_{k}^{(2,k)}\rangle$. Therefore, the probabilities read
     \begin{eqnarray}
\mathcal{P}_{s^{(1)},s'^{(2)}}=\mathcal{P}_{s^{(1)}}\mathcal{P}_{s'^{(2)}}\simeq \frac{1}{4},\label{mini1}
\end{eqnarray} and with the observer mind states:
\begin{eqnarray}
\mathcal{P}_{\spadesuit^{(1)},\spadesuit^{(2)}}=\mathcal{P}(\mathcal{O}_\uparrow^{(1)},\mathcal{O}_\uparrow^{(2)}),&
\mathcal{P}_{\spadesuit^{(1)},\heartsuit^{(2)}}=\mathcal{P}(\mathcal{O}_\uparrow^{(1)},\mathcal{O}_\downarrow^{(2)})
\nonumber \\
\mathcal{P}_{\heartsuit^{(1)},\spadesuit^{(2)}}=\mathcal{P}(\mathcal{O}_\downarrow^{(1)},\mathcal{O}_\uparrow^{(2)}),&
\mathcal{P}_{\heartsuit^{(1)},\heartsuit^{(2)}}=\mathcal{P}(\mathcal{O}_\downarrow^{(1)},\mathcal{O}_\downarrow^{(2)})\label{mini2}
\end{eqnarray} The previous formalism can be generalized to $N$ minds by introducing states like $|E_0,\mathcal{O}_0^{(1)},...,\mathcal{O}_0^{(N)}\rangle$ and using product spin-states $\bigotimes_{k=1}^{M}\bigotimes_{i=1}^{N}|s_{k}^{(i,k)}\rangle$. The unitary evolution together with Stosszahlansatz allow us to define the binomial probability 
\begin{eqnarray}
\mathcal{P}(N_\uparrow,N_\downarrow)=\frac{N!}{N_\uparrow !N_\downarrow !}\mathcal{P}(\mathcal{O}_\uparrow)^{N_\uparrow}\mathcal{P}(\mathcal{O}_\downarrow)^{N_\downarrow}\label{retrucmucb}.
\end{eqnarray} where $N_\uparrow$ and $N_\downarrow$ are the number of observer mind states $\mathcal{O}_\uparrow$ and $\mathcal{O}_\downarrow$ in the two branches $|\uparrow\rangle$ and $|\downarrow\rangle$ respectively~\footnote{We here used the notations $\mathcal{P}(\mathcal{O}_{\uparrow}^{(i)})=\mathcal{P}(\mathcal{O}_\uparrow)$ and $\mathcal{P}(\mathcal{O}_{\downarrow}^{(i})=\mathcal{P}(\mathcal{O}_\downarrow)$ and the fact that (from Eqs.~\ref{mini1},\ref{mini2}) we have $\mathcal{P}(\mathcal{O}_\alpha^{(i)},\mathcal{O}_\beta^{(j)})=\mathcal{P}(\mathcal{O}_\alpha^{(i)})\mathcal{P}(\mathcal{O}_\beta^{(j)})$ with $\alpha,\beta\in\{\uparrow,\downarrow\}$.}. This relation is obviously the same as Eq. \ref{retrucmuc} in the MMI of Albert and Loewer. Therefore, by a direct application of the law of large numbers we get:   
\begin{eqnarray}
\lim_{N \to +\infty} \frac{N_\uparrow}{N}=\mathcal{P}(\mathcal{O}_\uparrow)=\frac{1}{2}\nonumber \\
\lim_{N \to +\infty} \frac{N_\downarrow}{N}=\mathcal{P}(\mathcal{O}_\downarrow)=\frac{1}{2}
\label{drezet}
\end{eqnarray}  which shows that for typical configurations the minds are equiprobably distributed. Like in the MMI of Albert and Loewer the probability of a `mind-less hulk' is vanishing. Here the deductions is very similar to the one made by Albert and Loewer~\cite{AlbertLoewer1988,Albert1994} (see the discussion in Section \ref{sec2b}  after Eq.~\ref{retrucmubis} concerning the typicality of never meeting a mindless hulk).\\  
\begin{figure}[h]
\begin{center}
\includegraphics[width=8cm]{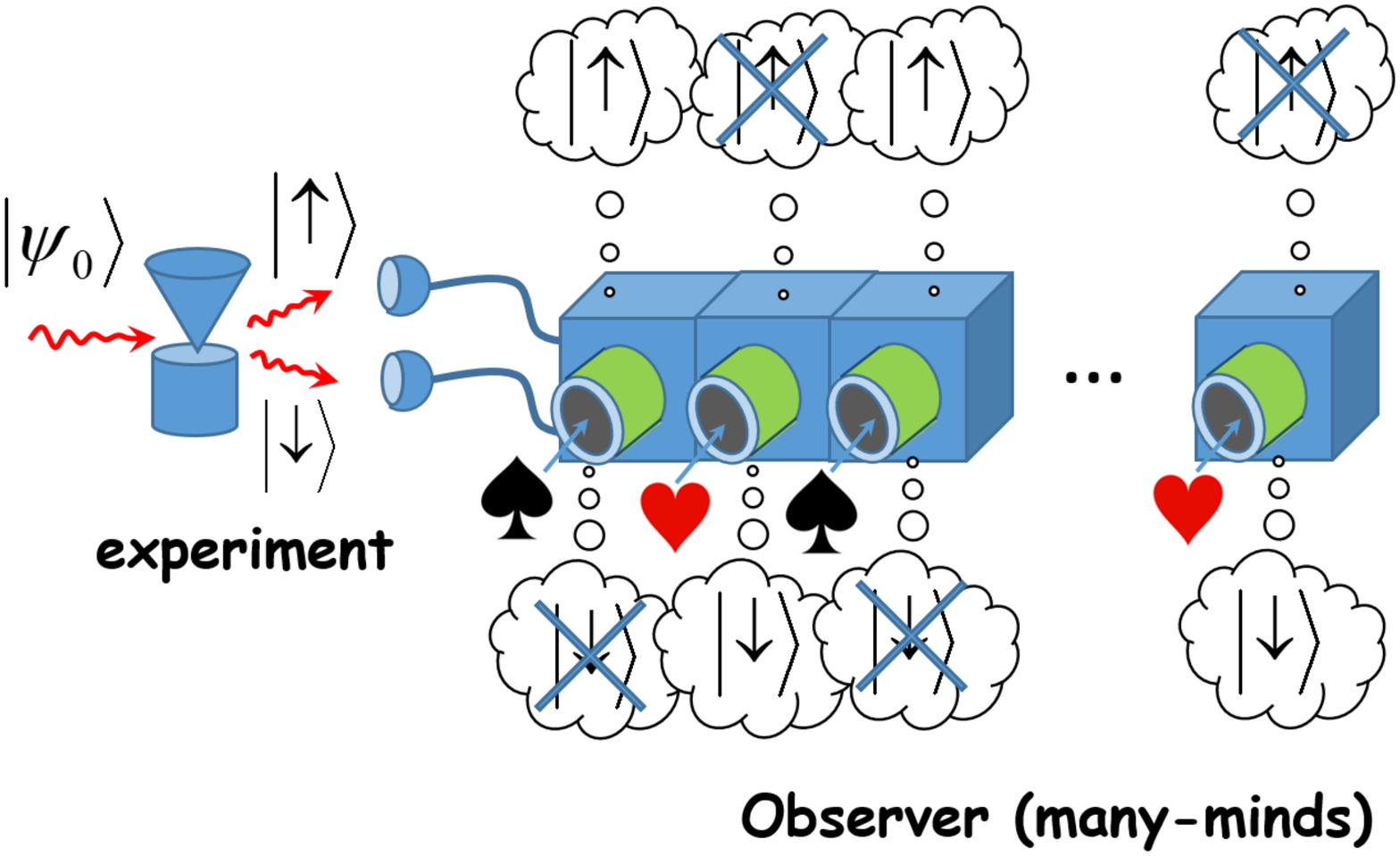} 
\caption{A quantum experiment involving an observer with several minds (compare with Fig.~\ref{figure1}). The observer is detecting a typical sequence of qubits $\spadesuit,\heartsuit,\spadesuit,...,\heartsuit$ prepared using the source (`oven') of Fig.~\ref{figure1} (a). This corresponds to a configuration for the different minds of the observer. } \label{figure2}
\end{center}
\end{figure}
\indent The previous narrative of Fig.~\ref{figure1} can be used to illustrate the idea. Again the source or oven prepares a typical sequence like $h_1:= [\spadesuit^{(1,1)},...,\heartsuit^{(N,1)}]$ in a large ensemble. This sequence is used by the quantum observer to drive the $N$ different minds in a `aware of the result' state or in a `not aware of a result' state (see Fig.~\ref{figure2}).  At the statistical level the number of minds observing the outcome  $\uparrow$ (or $\downarrow$) is approaching $N/2$ for large $N$ as explained before. This means that typically the probability of a maverick state including mindless hulk vanishes. Moreover, repeating again and again $M$ times the same experiment (involving typical sequences of qubits $h_k=[s_{k}^{(1,k)},...,s_{k}^{(N,k)}]$ with $k=1,...,M$) leads to Born's rule for each mind following a random but typical path.  \\
\indent In order to further generalize the previous procedure we can use the formal methodology of Zurek \cite{Zurek1998,Zurek2003a,Zurek2005}. For this we consider the time evolution $|\Psi_0\rangle\otimes |E_0\rangle \rightarrow \sqrt{\frac{1}{T}}\sum_{\alpha\in\Delta}|\alpha\rangle\otimes |E_\alpha\rangle $ for the symmetric Schmidt state ($T$ is an integer that defines the cardinality of $\Delta$). In presence of the observer and environment for a single mind we admit now instead of Eq. \ref{Drezetstate}: 
\begin{eqnarray}
|\Psi_0\rangle\otimes |E_0,\mathcal{O}_0^{(1)}\rangle\otimes |\tilde{\beta}^{(1)}\rangle\rightarrow \sqrt{\frac{1}{T}}
\sum_{\alpha\in\Delta, \alpha\neq\beta}|\alpha\rangle\otimes |E_\alpha,\emptyset^{(1)}\rangle \otimes |\tilde{\beta}^{(1)}\rangle\nonumber \\
+\sqrt{\frac{1}{T}}|\beta\rangle\otimes |E_\beta,\mathcal{O}_\beta^{(1)}\rangle \otimes |\tilde{\beta}^{(1)}\rangle
\label{Drezetstaten} 
\end{eqnarray}where we introduced a $T-$level quantum system $|\tilde{s}^{(1)}\rangle$ (i.e., with $s\in\Delta$) to drive the system. If $s=\beta$  the mind state $\mathcal{O}_0^{(1)}$ evolves into the $\mathcal{O}_\beta^{(1)}$ channel and the other channels are empty. Naturally, we have $T$ relations like Eq. \ref{Drezetstaten} corresponding to the $T$ different levels $|\tilde{s}^{(1)}\rangle$ and to the $T$ different branches. In complete analogy with the previous case we define  probabilities for the spin states and the related mind states $\mathcal{O}_\beta^{(i)}$ as:  
\begin{eqnarray}
\mathcal{P}_{\beta^{(i)}}=\mathcal{P}(\mathcal{O}_\beta^{(i)})\simeq \frac{1}{T}
\end{eqnarray}  We have from independence and the Stosszahlansatz for two mind states:
\begin{eqnarray}
\mathcal{P}_{\alpha^{(i)},\beta^{(j)}}=\mathcal{P}(\mathcal{O}_\alpha^{(i)},\mathcal{O}_\beta^{(i)})=\mathcal{P}_{\alpha^{(i)}}\mathcal{P}_{\beta^{(j)}}=\mathcal{P}(\mathcal{O}_\alpha^{(i)})\mathcal{P}(\mathcal{O}_\beta^{(j)})\simeq \frac{1}{T^2}
\end{eqnarray}  Finally, for a many-minds system (with $N$ mind states $\mathcal{O}^{(1)},...,\mathcal{O}^{(N)}$) we can define in analogy with Eq. \ref{retrucmucb} the multinomial probability for having the distribution $\{N_\alpha\}$ where $N_\alpha$ is the number of mind states $\mathcal{O}_\alpha$ in the branch $|\alpha\rangle$ (with the constraint $\sum_{\alpha\in\Delta}N_\alpha=N$):  
\begin{eqnarray}
\mathcal{P}(\{N_\alpha\})=\frac{N!}{\Pi_\alpha N_\alpha!} \Pi_\alpha\mathcal{P}(\mathcal{O}_\alpha)^{N_\alpha}\simeq \frac{N!}{\Pi_\alpha N_\alpha!}\frac{1}{T^N}. \label{drezetA}\end{eqnarray}
In the typical regime we deduce from the law of large numbers the relation 
\begin{eqnarray}
\lim_{N \to +\infty}\frac{N_\alpha}{N}=\mathcal{P}(\mathcal{O}_\alpha)=\frac{1}{T}
\label{drezetbb}
\end{eqnarray}  which shows that with  the state $\sqrt{\frac{1}{T}}\sum_{\alpha\in\Delta}|\alpha\rangle$ and for typical configurations the minds are equally distributed  between the branches (i.e., the mindlless hulk problem is again avoided). \\ 
\indent The last step of our deduction follows strictly the fine graining trick of Zurek discussed in Section \ref{sec2d}. Indeed, starting with a quantum state $|\Psi_0\rangle=\sum_{a}\sqrt{\mathcal{P}_a}|a\rangle$ with $\mathcal{P}_a=\frac{T_a}{T}$ a rational number we formally repeat the steps from  Eq.~\ref{Zurek7} to Eq. \ref{Zurek10} and  obtain after entanglement with a specifically designed environment the Schmidt state $|\Psi_0\rangle\otimes |E_0\rangle \rightarrow \sqrt{\frac{1}{T}}\sum_{\alpha\in\Delta}|\alpha\rangle\otimes |E_\alpha\rangle $.     The rest of the reasoning is similar to the one leading to Eq. \ref{drezetA} and Eq. \ref{drezetbb}: we deduce the probability 
\begin{eqnarray}
\lim_{N \to +\infty} \sum_{\alpha\in\Delta_a}\frac{N_\alpha}{N}=\sum_{\alpha\in\Delta_a}\mathcal{P}(\mathcal{O}_\alpha)=\frac{T_a}{T}:=\mathcal{P}_a
\label{drezetc}
\end{eqnarray} where $\Delta_a$ is the subset of $\Delta$ associated with the quantum state $|a\rangle$ (see Eq.~\ref{Zurek8}). This closes our derivation of Born's rule. We stress that if the formal deduction is similar to the one made by Zurek with his coarse-graining trick we don'tuse envariance as a fundamenal property here. Indeed, invariance is seen by Zurek as a quantum version of the Keynes-Laplace indifference principle.  However, here we want to ground the MMI (and thus the MWI) on a objective definition of probability based on molecular chaos  on the initial conditions of quantum systems (like our qubits $\spadesuit/\heartsuit$). This is a strategy used in classical statistical mechanical (or in the PWI) that don't require subjective ignorance at a fundamental level (even  if ignorance can play a role in the application of the probability calculus).\\ 
\indent We emphasize, that the role of the environment is central since it allows us to define unambiguously decohered Worlds evolving independently. Therefore, the theory agrees for all practical purposes with the standard  quantum mechanical interpretation of decoherence. Like in Zurek existential interpretation \cite{Zurek1998,Zurek2003b} the unitary evolution is all what is required for the theory to hold. However, here Born's rule results as a contingent  consequence of the dynamic relying on the Stosszahlansatz hypothesis. Like in Everett's (but unlike in Zurek's) work the role of the observer interacting with the gas of $T-$level systems is here central.  Also,  we stress that the theory we propose here is naturally generalized to systems of several observers with huge numbers of minds $N\rightarrow +\infty $. This condition is mandatory  as stressed by Albert and Loewer in order to recover a common experience agreement between the separate perspective of the various observers (i.e., in order to avoid the mindless Hulk dilemma discussed in \cite{AlbertLoewer1988,Albert1994}).      
Finally, we emphasize once more that the model is fully unitary and doesn't require a mind/brain dualism (i.e., unlike some readings of the original MMI \cite{AlbertLoewer1988}).  Here, the minds are physically linked to the brain and define some quantum excitations of a mechanical structure.
\section{Conclusion and comments}\label{secconc}
\indent To conclude this work several remarks are  necessary~\footnote{An intereresting issue that we don't consider in this article (for reasons of space) concerns Bell's theorem  and the notion of non-locality~\cite{Bell2004,Bricmont2016,Maudlin2019} studied in connections with the MWI \cite{Tipler2014} and the MMI \cite{AlbertLoewer1988,Albert1994}.  }. First, observe that the present derivation is strongly related to Zurek's work about the Laplace indifference principle.  Going back to Eq. \ref{Drezetstaten1} and Eq. \ref{Drezetstaten2} we see that the main difference concern the permutation between two observer mind states $\mathcal{O}^{(1)}$ and $\mathcal{O}^{(2)}$ in the two branches corresponding to the observable $|\uparrow\rangle$ or $|\downarrow\rangle$. Removing the irrelevant degrees of freedom we see that at the end of the unitary evolution we get either  
\begin{eqnarray}
\sqrt{\frac{1}{2}}\left( |\uparrow\rangle\otimes |E_\uparrow,\mathcal{O}_\uparrow^{(1)},\emptyset^{(2)}\rangle+|\downarrow\rangle\otimes |E_\downarrow,\emptyset^{(1)},\mathcal{O}_\downarrow^{(2)}\rangle\right), 
\label{Drezetstaten1B}
\end{eqnarray}    
or 
\begin{eqnarray}
\sqrt{\frac{1}{2}}\left( |\uparrow\rangle\otimes |E_\uparrow,\emptyset^{(1)},\mathcal{O}_\uparrow^{(2)}\rangle+|\downarrow\rangle\otimes |E_\downarrow,\mathcal{O}_\downarrow^{(1)},\emptyset^{(2)}\rangle\right) ,
\label{Drezetstaten2B}
\end{eqnarray}  depending which state (i.e., $|\spadesuit^{(1)}\rangle\otimes |\heartsuit^{(2)}\rangle$ or $|\heartsuit^{(1)}\rangle\otimes |\spadesuit^{(2)}\rangle$) the Nature `randomly' assigned to the full system. Now, the objective probability (i.e., relative frequency) of each evolution is typically $\frac{1}{2}$ in agreement with our previous discussion of the Stosszahlansatz. Therefore, we get 
\begin{eqnarray}
\mathcal{P}(\mathcal{O}_\uparrow^{(i)})=\mathcal{P}(\mathcal{O}_\downarrow^{(i)})=\frac{1}{2}
\label{DrezetEND}
\end{eqnarray} 
for $i=1,2$. There is complete indifference for the observer mind state $i$ about where he will finish his journey. This conveys the spirit of Laplace indifference principle which is here driven by symmetry like for a classical die tossing. On the one side, this  clean self-locating uncertainty for each  mind state is completely classical since it is driven by statistical distributions associated with a  Stosszahlansatz for the heart and spade permutations. On the other side, this result is fully quantum and unitary. Unlike in the original  MMI of Albert and Loewer no genuine stochastic process breaking the unitarity of the quantum evolution has to be invoked.  Compared to the self-locating discussion given by Zurek, Deutsch, Wallace, and Carroll and Sebens our procedure doesn't suffer from the incoherence problem associated with the standard MWI.  Clearly, this means that we dont try to stick rigorously to the usual debate done by usual proponents of the MWI (who probably will not agree with our solution~\footnote{In the usual MWI the incoherence problem explicitely excludes initial conditions uncertainty and molecular chaos as a source of randomness. This is not the case in our MMI where we introduce ingredients of statistical mechanics often used in hidden-variable theories like the PWI.}).  The price to be paid is of course high since we need to define a unitary version of the MMI and thus have to introduce several mind states to avoid the schizophrenic quantum superposition of the Schr\"odinger cat. \\ 
\indent The previous analysis prompts at least two potential criticisms. The first criticism concerns the causal structure of the model used in that work which is very conspiratorial. Indeed, the model used by Zurek is already conspiratorial since in order to recover Born's rule for a general quantum state  we must include a fine graining procedure (which we named a trick) which looks physically superdeterministic.  In this approach  the fine graining procedure can be experimentally implemented by using logical gates and beam splitters (see for instance the discussions and proposals made by Vaidman \cite{Vaidman2020}). The idea is to introduce states like $|a^{(S)}\rangle=\frac{1}{\sqrt{N_a}}\sum_{\alpha\in\Delta_a}|\alpha^{(S)}\rangle$ to transform a general wave-function given by Eq. \ref{Zurek7} into a symmetric Schmidt state as given by Eq. \ref{Zurek9}. However, this fine graining is necessarily wave-function dependent and by changing the probability coeeficents in the initial state we would have to modify completely the fine grains for the future experiments. Therefore, from the point of view of causality where the observer is selecting the beam splitters and other apparatus this looks like the fine grains was decided in advance in a conspiratorial way for reproducing Born's rule for a very specific problem. This is of course not necessarily a fatal objection if we are ready to accept such features. Many other interpretations of quantum mechanics involve superdeterministic or even retrocausal properties but we have at least to be aware of the problem. \\ 
\indent The second criticism concerns of course the notion of minds introduced in the present work.  We fully agree that this notion is  speculative. In fact, Albert recently presented negative comments concerning the value of this whole business about many minds and about his own work done with Loewer. Albert  wrote: 
\begin{quote}
\textit{The many-minds interpretation of quantum mechanics that Barry Loewer and I discussed twenty-five or so years ago, and which is rehearsed in chapter of \cite{Albert1994}, was a (bad, silly, tasteless, hopless, explicitely dualist) attempts of coming to terms with that realization.}~\cite{Albert2015}, p.~163. 
\end{quote} 
The model discussed here is presented as a solution for making sense of the MMI and MWI or may be should we say for saving the whole MWI enterprise. Yet, several authors got interested in the past years about the possible role of minds and consciousness in the interpretations of quantum mechanics. So, may be this is a good strategy to try. The `toy' model considered here is far from being complete. However, our model is fully deterministic and preserves unitarity together with the core ideas of the functionalist approach to the mind/body problem. It preserves all the advantages of the original MMI without its bad features such as dualism and stochasticity.  Moreover, our approach is physically appealing from the statistical point of view since it relies on the `typical' distribution of qubits $\heartsuit,\spadesuit,...$ for driving the various minds. Therefore, like in classical and Bohmian statistical mechanics this randomness is associated with a typical choice of initial conditions for our Universe. This shows that in order to recover Born's rule or the Born-Vaidman postulate in the fully unitary MWI and MMI we have to introduce random elements associated with the cosmological initial conditions. Like in classical or Bohmian mechanics these initial conditions are contingent elements which must be added to the theory in order to justify the statiscal observations. In turn, we  hope that the model suggested here will open new perspectives which could motivate further work in that fascinating area.\\



\begin{thebibliography}{99}
\bibitem{Everett1957}
Everett, H. III.: `Relative State'  formulation of quantum mechanics. Rev. Mod. Phys. \textbf{29}, 454 (1957).
\bibitem{Barrett2012}
Barett, J. A. and Byrne, P.: The Everett interpretation of quantum mechanics: Collected works 1955-1980 with commentary. Princeton University Press, Princeton (2012).
\bibitem{DeWitt1973}
DeWitt, B. S. and Graham, N.: The Many-Worlds interpretation of quantum mechanics. Princeton University Press, Princeton (1973).
\bibitem{AlbertLoewer1988}
Albert, D. and Loewer, B.: Interpreting the many-worlds intepretation. Synthese \textbf{77}, 195 (1988).
\bibitem{Albert1994}
Albert, Z. D.: Quantum mechanics and experience.  Harvard University Press, Harvard (1994).
\bibitem{Lockwood1989}
Lockwood, M.: Mind, brain and the quantum: the compound `I'. Blackwell Publishers, Oxford (1989).
\bibitem{Lockwood1996a}
Lockwood, M.: `Many Minds' interpretations of quantum mechanics. Brit. J. Phil. Sci. \textbf{47}, 159 (1996).
\bibitem{Donald1}
Donald, M.~J.: Quantum theory and the brain. Proc. Roy. Soc. A \textbf{427}, 43 (1990).
\bibitem{Donald2}
Donald, M.~J.: A priori probability and localized observers. Found. Phys. \textbf{22}, 1111 (1992).
\bibitem{Donald3}
Donald, M.~J.: On many-minds interpretations of quantum theory. arXiv:quant-ph/9703008 (1997).
\bibitem{Deutsch1999}
Deutsch, D.: Quantum  theory of probability and decisions. Proc. R. Soc. A \textbf{455}, 3129 (1999).
\bibitem{Wallace2012}
Wallace, D.: The emergent multiverse: quantum theory according to the Everett interpretation. Oxford University Press, Oxford (2012).
\bibitem{Zurek2005}
Zurek, W. H .: Probabilities from entanglement, Born's rule $p_k=|\psi_k|^2$ from envariance. Phys. Rev. A \textbf{71}, 052105 (2005).
\bibitem{Sebens2016}
Sebens, C. T., and Carroll, S.: Self-locating uncertainty and the origin of probability in Everettian quantum mechanics. Brit. J. Phil. Sc. \textbf{69}, 25 (2018).
\bibitem{McQueen}
McQueen, K.~J. Vaidman, L.: In defence of the self-location uncertainty account of probability in the many-worlds interpretation. Studies in  History and Philosophy of Science part B.: Studies in History and Philosophy of Modern Physics  \textbf{66}, 14-23(2019).
\bibitem{Gleason1957}
Gleason, A. M.: Measures on the closed subspaces of a Hilbert space. Indiana Univ. Math. J. \textbf{6}, 885-897 (1957).
\bibitem{Lubkin1979}
Lubkin, E.: An application of ideal experiments to quantum mechanical measurement theory. Int. J. Phys. \textbf{18}, 165-177 (1979).
\bibitem{Goldstein2012}
Goldstein, S.: Typicality and notions of probability, in Probability in physics, edited by  Y. ben-Menahem and M. Hemmo, The Frontiers Collection. Springer, Berlin (2012).
\bibitem{Durr1992}
D\"{u}rr, D., Goldstein, S., Zangh\`{i}, N.:Quantum equilibrium and the origin of absolute uncertainty. J.~Stat.~Phys. \textbf{67}, 843 (1992).
\bibitem{Hartle1968}
Hartle, J. B.: Quantum mechanics of individual systems. Am.~J.~Phys. \textbf{36}, 704 (1968).
\bibitem{DeWitt1971}
DeWitt, B. S.: Quantum mechanics and reality. Phys. Today \textbf{23}, 30 (1970).
\bibitem{Farhi1989}
Farhi, E. and Goldstone, J.: How probability arises in quantum mechanics. Ann. Phys. \textbf{192}, 368 (1989).
\bibitem{Aharonov2002}
Aharonov, Y. and Reznik, B.: How macroscopic properties dictate microscopic probabilities. Phys. Rev. A \textbf{65}, 052116 (2002).
\bibitem{GellMann}
GellMann, M. and Hartle, J. B.: Quantum Mechanics in
the Light of Quantum Cosmology, in Complexity, Entropy,
and the Physics of Information, SFI Studies in
the Sciences of Complexity, Vol. VIII, edited by W. Zurek. Addison Wesley, Reading, MA(1990).
\bibitem{Ballentine1973}
Ballentine, L. E.: Can the statistical postulate of quantum theory be derived?—A critique of the many-universes interpretation. Found. Phys. \textbf{3}, 229 (1973).
\bibitem{Kent1990}
Kent, A.: Against many-worlds interpretation. Int. J. Mod. Phys. A \textbf{5}, 1745 (1990).
\bibitem{Squires1990}
Squires, E.: On an alleged `proof' of the quantum probability law. Phys. Lett.  A \textbf{145}, 67 (1990).
\bibitem{Aguire2011}
Aguire, A. and Tegmark, M.: Born in an infinite universe: A cosmological interpretation of quantum mechanics. Phys. Rev. D \textbf{84}, 105002 (2011).
\bibitem{Wallacevideo}
Wallace, D.: The Probability puzzle and many-worlds interpretation of quantum mechanics. https://www.youtube.com/watch?v=8turL6Xnf9U  (2015).
\bibitem{deBroglie} Bacciagaluppi, G. and Valentini, A.: Quantum theory at the crossroads: Reconsidering the 1927 Solvay Conference. Cambridge University Press, Cambridge (2009).  
\bibitem{BohmHiley} Bohm., D and Hiley, B.J.: The undivided Universe. Routledge, London, (1993).
\bibitem{Pauli}
Pauli, W.: Remarques sur le probl\`{e}me des param\`{e}tres cach\'{e}s dans la m\'{e}canique quantique et sur la th\'{e}orie de l'onde pilote, in Louis de Broglie Physicien et Penseur, pp. 33-42. Albin Michel, Paris (1953).
\bibitem{Valentini}
Valentini, A.: On the pilot-wave theory of classical, quantum and subquantum physics. International School for advanced studies, Trieste (1992).
\bibitem{DrezetIJQF}
Drezet, A.: Analysis of Everett’s quantum interpretation from the point of view of a Bohmian. Int. J. Quant. Found. \textbf{2}, 1-22 (2016).
\bibitem{Bricmont2016}
Bricmont, J.: Making sense of quantum mechanics. Springer, Berlin (2016).
\bibitem{Albert2015}
Albert, Z. D.: After physics.  Harvard University Press, Harvard (2015).
\bibitem{Barrett2019}
Barrett, J. A.: Typical worlds. arXiv:1912.05312v1 (2019).
\bibitem{Vaidman1998}
Vaidman, L.: On schizophrenic experiences of the neutron or why we should belive in the many-worlds interpretation of quantum mechanics. International Studies in the Philosophy of Science \textbf{12}, 245 (1998).
\bibitem{Vaidman2012}
Vaidman, L.: Probability in the many-worlds interpretation of quantum mechanics, in Probability in physics, The frontiers collection, edited by Y. Ben Menuhem, M. Hemmo. Springer, Berlin (2012).
\bibitem{Vaidman2014}
Vaidman, L.: Quantum theory and determinism. Quantum Studies: Mathematics and Foundations \textbf{1}, 5 (2014).
\bibitem{Vaidman2020}
Vaidman, L.: Derivations of the Born rule, in Quantum, probability, logic in physics, Jerusalem Studies in philosophy and history of science, edited by M. Hemmo, O. Shenker. Springer Nature, Switzerland (2020).
\bibitem{Tappenden2010}
Tappenden, P.: Evidence and uncertainty in Everett's multiverse. Brit. J. Phil. Sci. \textbf{62}, 99 (2010).
\bibitem{Tappenden2019}
Tappenden, P.: Everettian theory as a pure wave mechanics plus a no-collapse probability postulate. Synthese  https://doi.org/10.1007/s11229-019-02467-4 (2019).
\bibitem{Greaves2004}
Greaves, H.: Understanding Deutsch's probability in a deterministic multiverse. Studies in  History and Philosophy of Science part B.: Studies in History and Philosophy of Modern Physics \textbf{35}, 423 (2004).
\bibitem{Maudlin2019}
Maudlin, T.: Philosophy of Physics: Quantum Theory.  Princeton University Press, Princeton (2019).
\bibitem{bookMWI} Saunders., S. Barrett, J. Kent, A.  and Wallace, D.: Many Worlds?: Everett, quantum theory, and Reality. Oxford University Press, Oxford, (2010).
\bibitem{Bell2004}
Bell, J.S.: Speakable and unspeakable in quantum mechanics, second edition. Cambridge University Press, Cambridge (2004).
\bibitem{Zeh}
Zeh, H. D.: arXiv:quant-ph/9908084v3 (2000).
\bibitem{vonNeumann}
von Neumann, J.: Mathematical foundations of quantum mechanics. Princeton University Press, Princeton (1955).
\bibitem{LondonBauer}
London, F. Bauer., E.: The Theory of Observation in Quantum Mechanics in Wheeler J. A. and Zurek W. H. Quantum Theory and Measurement, p. 217. Princeton University Press, Princeton (1983).
\bibitem{Wigner}
Wigner, E.: Remarks on the mind-body question, in the scientist speculates, edited by Good, I.~J. Basic Books, New York (1962).
\bibitem{Barrett1995}
Barrett, J.~A.:The single-mind and many-minds versions of quantum mechanics.  Erkenntnis \textbf{42}, 89 (1995).
\bibitem{Brown1996}
Brown, H.~R.: Mindful of quantum possibilities. Brit. J. Phil. Sci. \textbf{47}, 189 (1996).
\bibitem{Butterfield1996}
Butterfield, J.: Whither the minds. Brit. J. Phil. Sci. \textbf{47}, 200 (1996).
\bibitem{Deutsch1996}
Deutsch, D.: Comment on Lockwood. Brit. J. Phil. Sci. \textbf{47}, 222 (1996).
\bibitem{Loewer1996}
Loewer, B.: Comment on Lockwood. Brit. J. Phil. Sci. \textbf{47}, 229 (1996).
\bibitem{Saunders1996}
Saunders, S.: Comment on Lockwood. Brit. J. Phil. Sci. \textbf{47}, 241 (1996).
\bibitem{Papineau1996}
Papineau, D.: Many Minds are no worse than one. Brit. J. Phil. Sci. \textbf{47}, 233 (1996).
\bibitem{Lockwood1996b}
Lockwood, M.: `Many Minds' interpretations of quantum mechanics: replies to replies. Brit. J. Phil. Sci. \textbf{47}, 445 (1996).
\bibitem{Papineau}
Papineau, D.: Many Minds and probabilities. Analysis. \textbf{55}, 239 (1995).
\bibitem{Barnum}
Barnum, H. Caves, C. M., Finkelstein J., Fuchs, C.~A. Schack, R.: Quantum probability from decision theory? Proc. Roy. Soc. A \textbf{456}, 1175 (2000).
\bibitem{Pitowsky}
Hemmo, M. Pitowsky, I.: Quantum probability and many worlds. Studies in  History and Philosophy of Science part B.: Studies in History and Philosophy of Modern Physics \textbf{38}, 333 (2007).
\bibitem{Wallace2003a}
Wallace, D.: Everett and structure. Studies in  History and Philosophy of Science part B.: Studies in History and Philosophy of Modern Physics \textbf{34}, 87 (2003).
\bibitem{Wallace2003b}
Wallace, D.: Everettian rationality: defending Deutsch's approach to probability in the Everett interpretation. Studies in  History and Philosophy of Science part B.: Studies in History and Philosophy of Modern Physics \textbf{34}, 415 (2003).
\bibitem{Wallace2007}
Wallace, D.: Quantum  probability from subjective likelihood. Studies in  History and Philosophy of Science part B.: Studies in History and Philosophy of Modern Physics \textbf{38}, 311 (2007).
\bibitem{Saunders2005}
Saunders, S.: What is a probability?, in  Quo Vadis quantum mechanics. edited by Eltizur, A. Dolev, S., and Kolenda N. Springer, Berlin (2005).
\bibitem{Saunders2008}
Saunders, S. Wallace, D.: Branching and uncertainty. Brit. J. Phil. Sci. \textbf{59}, 293 (2008).
\bibitem{Zurek2003a}
Zurek, W.~H: Environment-assisted invariance, entanglement, and probabilities in quantum physics.  Phys. Rev. Lett. \textbf{90}, 120404 (2003).
\bibitem{Zurek2003b}
Zurek, W.~H: Decoherence, einselection, and the quantum origins of the classical.  Rev. Mod. Phys. \textbf{75}, 715 (2003).
\bibitem{Zurek2014}
Zurek, W.~H: Quantum Darwinism, classical reality, and the randomness of quantum jumps. Phys. Today \textbf{67}, 44 (2014).
\bibitem{Barnum2}
Barnum, H.: No-signalling-based version of Zurek's derivation of quantum probabilities: A note on `Environment-assisted invariance, entanglement, and probabilities in quantum physics'. arXiv:quant-ph/0312150v1 (2003).
\bibitem{Zurek1998}
Zurek, W.~H: Decoherence, einselection and the existential interpretation (the rough guide). Proc. Roy. Soc. A \textbf{356}, 1793 (1998).
\bibitem{Deutsch1985}
Deutsch, D.: Quantum theory as a universal physical theory. Int. J. Th. Phys. \textbf{24}, 1 (1985).
\bibitem{Tipler2014}
Tipler, F.: Quantum nonlocality does not exist. PNAS \textbf{111}, 11281 (2014).
\bibitem{Bostrom2014}
Bostr\"{o}m, K.~J.: Quantum mechanics as a deterministic theory of a continuum of worlds. Quantum Studies: Mathematics and Foundations  \textbf{2}, 315 (2014).
\bibitem{Sebens2014}
Sebens, C.~T.: Quantum mechanics as classical physics. Philosophy of Science  \textbf{82}, 266 (2014).
\bibitem{Hall2014}
Hall, M.~J.~W. Deckert, D.-A. Wiseman, H.~M.: Quantum phenomena modeled by interactions between many classical worlds. Phys. Rev. X  \textbf{4}, 041013 (2014).
\bibitem{Kent2015}
Kent, A.: Does it makes sense to speak of self-locating uncertainty in the universal  wave-function? Remarks on Sebens and Carroll. Found. Phys. \textbf{45}, 211 (2015).
\bibitem{Drezet}
Drezet, A.: How to justify Born’s rule using the pilot wave theory of de Broglie?. Ann. Found. Louis de Broglie \textbf{42}, 103 (2017).
\end{thebibliography}
\end{document}